\documentclass{aa}  
\AtBeginDocument{\let\linenumbers\nolinenumbers}
\usepackage[utf8]{inputenc}
\usepackage{booktabs}
\usepackage[varg]{txfonts}
\usepackage{graphicx}
\usepackage{amsmath}
\usepackage{amssymb}
\usepackage{xcolor}
\usepackage{natbib}
\bibpunct{(}{)}{;}{a}{}{,}
\usepackage{url}
\usepackage[colorlinks=true]{hyperref}

\hypersetup{colorlinks=true,allcolors=black,citecolor=blue}
\usepackage[
    separate-uncertainty=true,
    multi-part-units=single,
    range-phrase=--,
    range-units=single
]{siunitx}
\newcommand{\kms}{\,km\,s$^{-1}$}

\newcommand{\okina}{\textquoteleft} 

\authorrunning{Artigau et al.}

\usepackage{tabularx}

\def\welo{\hbox{Welopō\okina ai}}

\def\tyc{Welopō\okina ai\,Aa}
\def\innerwd{Welopō\okina ai\,Ab}
\def\outerwd{Welopō\okina ai\,B}
\def\GtwoZeroThree{G~203-47}

\def\Bl{\hbox{$B_{\rm \ell}$}}
\def\chisqr{\hbox{$\chi^2_{\rm r}$}}
\newcommand{\AlopekeH}{\okina Alopeke} 
\defcitealias{Stefanov2025a}{S25}

\makeatletter
\DeclareRobustCommand
  \shortvdots{\vbox{\baselineskip4\p@ \lineskiplimit\z@
    \hbox{.}\hbox{.}\hbox{.}}}
\makeatother 

\DeclareSIUnit{\au}{au}
\DeclareSIUnit{\hjulianday}{HJD}
\DeclareSIUnit{\julianday}{BJD}
\DeclareSIUnit{\mas}{mas}
\DeclareSIUnit{\mag}{mag}
\DeclareSIUnit{\parsec}{pc}
\DeclareSIUnit{\pixel}{px}
\DeclareSIUnit{\dex}{dex}
\DeclareSIUnit{\solarmass}{M\textsubscript{\(\odot\)}}
\DeclareSIUnit{\solarradius}{R\textsubscript{\(\odot\)}}
\DeclareSIUnit{\solarluminosity}{L\textsubscript{\(\odot\)}}
\DeclareSIUnit{\earthmass}{M\textsubscript{\(\oplus\)}}
\DeclareSIUnit{\earthflux}{\text{$\Phi$}\textsubscript{\(\oplus\)}}
\DeclareSIUnit{\year}{yr}
\newcommand{\gauss}{G}

\begin{document}

\renewcommand{\today}{\number\day\space\ifcase\month\or January\or February\or March\or
  April\or May\or June\or July\or August\or September\or October\or November\or December\fi
  \space\number\year}

\title{Welopō\okina ai: A peculiar triple system 8\,pc from the Sun}
\author{\'Etienne Artigau\inst{1,2}\corrauth{etienne.artigau@umontreal.ca}
\and Charles-Antoine Parent\inst{3,2,1}\email{charles-antoine.parent@umontreal.ca}
\and Atanas K. Stefanov\inst{4,5}\email{atanas.stefanov@iac.es}
\and \'Erika Le Bourdais\inst{3,2,1}\email{erika.le.bourdais@umontreal.ca}
\and Souleymane~Ouedraogo\inst{6,7,1}\email{souley.ouedraogo@ujkz.bf}
\and Alexandrine L'Heureux\inst{1}\email{alexandrine.lheureux@umontreal.ca}
\and Patrick Dufour\inst{3,2,1}\email{patrick.dufour@umontreal.ca}
\and Ren\'e Doyon\inst{1,2}\email{rene.doyon@umontreal.ca}
\and Neil J. Cook\inst{1}\email{neil.cook@umontreal.ca}
\and Charles~Cadieux\inst{8,1}\email{charles.cadieux@unige.ch}
\and Cl\'emence Fontanive\inst{9,1,2}\email{clemence.fontanive@roe.ac.uk}
\and Luc Arnold\inst{10}\email{larnold@cfht.hawaii.edu}
\and Jean-Fran\c{c}ois Donati\inst{11}\email{jean-francois.donati@irap.omp.eu}
\and Matthias R. Schreiber\inst{12}\email{Matthias.Schreiber@usm.cl}
\and Ryan~Cloutier\inst{13}\email{ryan.cloutier@mcmaster.ca}
\and Xavier~Delfosse\inst{14}\email{Xavier.Delfosse@univ-grenoble-alpes.fr}
\and C\'eline Reyl\'e\inst{15}\email{celine@obs-besancon.fr}
\and Thierry Forveille\inst{14}\email{Thierry.Forveille@univ-grenoble-alpes.fr}
\and Tianqi Cang\inst{16}\email{cangtq@mail.bnu.edu.cn}
\and Claire Moutou\inst{11}\email{claire.moutou@irap.omp.eu}
\and Alexandre~C.~M.~Correia\inst{17,18}\email{acor@uc.pt}}
\institute{
\inst{1}Institut Trottier de recherche sur les exoplan\`etes, D\'epartement de Physique, Universit\'e de Montr\'eal, Montr\'eal, Qu\'ebec, Canada\\
\inst{2}Observatoire du Mont-M\'egantic, Qu\'ebec, Canada\\
\inst{3}D\'epartement de Physique, Universit\'e de Montr\'eal, Montr\'eal, Qu\'ebec, Canada\\
\inst{4}Instituto de Astrof\'isica de Canarias (IAC), Calle V\'ia L\'actea s/n, 38205 La Laguna, Tenerife, Spain\\
\inst{5}Departamento de Astrof\'isica, Universidad de La Laguna (ULL), 38206 La Laguna, Tenerife, Spain\\
\inst{6}Laboratoire de Physique et de Chimie de l’Environnement, Universit\'e Joseph Ki-Zerbo, 03 BP 7021 Ouagadougou 03, Burkina Faso\\
\inst{7}Universit\'e C\^ote d'Azur, Observatoire de la C\^ote d'Azur, Centre National de la Recherche Scientifique (CNRS), Laboratoire Lagrange, France\\
\inst{8}Observatoire de Gen\`eve, D\'epartement d’Astronomie, Universit\'e de Gen\`eve, Chemin Pegasi 51, 1290 Versoix, Switzerland\\
\inst{9}SUPA (Scottish Universities Physics Alliance), Institute for Astronomy, University of Edinburgh, Blackford Hill, Edinburgh EH9 3HJ, UK\\
\inst{10}CNRS UAR2208 - Canada France Hawai‘i Telescope Corporation (CFHT), 65-1238 Mamalahoa Hwy, Kamuela, HI 96743, USA\\
\inst{11}CNRS, OMP, Universit\'e de Toulouse, 14 Avenue Belin, F-31400 Toulouse, France\\
\inst{12}Departamento de F\'isica, Universidad T\'ecnica Federico Santa Mar\'ia, Av. Espa\~na 1680, Valpara\'iso, Chile\\
\inst{13}Department of Physics \& Astronomy, McMaster University, 1280 Main St W, Hamilton, ON, L8S 4L8, Canada\\
\inst{14}Univ. Grenoble Alpes, CNRS, IPAG, 38000 Grenoble, France\\
\inst{15}Universit\'e Marie et Louis Pasteur, CNRS, Observatoire des Sciences de l'Univers THETA Franche-Comt\'e Bourgogne, Institut UTINAM, Observatoire de Besan\c{c}on, BP 1615, 25010 Besan\c{c}on Cedex, France\\
\inst{16}School of Physics and Astronomy, Beijing Normal University, No.19, Xinjiekouwai Street, Beijing, 100875, People's Republic of China\\
\inst{17}CFisUC (Centro de F\'isica da Universidade de Coimbra), Departamento de F\'isica, Universidade de Coimbra, 3004-516 Coimbra, Portugal\\
\inst{18}LTE (Laboratoire Temps-Espace), Observatoire de Paris, Universit\'e PSL, Sorbonne, CNRS, 75014 Paris, France}

\date{Received 31 May 2026 / Accepted dd September 2026}

\def\sysDistance{\SI{8.46}{\parsec}}

\def\AaParallaxDRtwo{\SI{123.06 \pm 0.59}{\mas}}

\def\AaAstrometricChiSq{$1.98\times10^{5}$}

\def\AaGmag{\SI{9.409 \pm 0.010}{\mag}}

\def\sysDistanceFull{$8.4635\pm0.0011$}

\def\AaGaiaDRthreeID{2202703050388170880}

\def\BGaiaDRthreeID{2202703050401536000}

\def\AaRA{21:51:38.153}

\def\AaDec{+59:17:40.01}

\def\BRA{21:51:39.927}

\def\BDec{+59:17:34.55}

\def\AaMuAlpha{$-79.2\pm1.2$}

\def\AaMuDelta{$10.5\pm1.2$}

\def\BMuAlpha{$-86.742\pm0.018$}

\def\BMuDelta{$-19.237\pm0.018$}

\def\BParallaxGDRthree{$118.155\pm0.016$}

\def\AaGBPmag{$10.842\pm0.010$}

\def\AaGRPmag{$8.199\pm0.010$}

\def\BGBPmag{$14.814\pm0.010$}

\def\BGRPmag{$13.763\pm0.010$}

\def\AbFiveSixTwomag{$15.01\pm0.38$}

\def\AaFiveSixTwomag{10.382}

\def\AbEightThreeTwomag{$14.27\pm0.27$}

\def\AaEightThreeTwomag{8.007}

\def\TycJmag{$6.529\pm0.025$}

\def\TycHmag{$5.860\pm0.038$}

\def\TycKmag{$5.651\pm0.020$}

\def\nRVframes{411}

\def\nRVepochs{103}

\def\durationRv{\SI{4.8}{\year}}

\def\spirouMedianSNR{103}

\def\spirouMedianRVerr{\SI{3.0}{\metre\per\second}}
\def\spirouMedianNightlyRVerr{\SI{1.5}{\metre\per\second}}

\def\rvPeakToPeak{\SI{16}{\kilo\metre\per\second}}

\def\aperoVersion{0.7.292}

\def\templateSNR{$\sim$2000}

\def\nStokesVtotal{104}

\def\nStokesVkept{98}

\def\BlRange{$-17$ to $+17$\,G}

\def\BlMedian{\SI{2}{\gauss}}

\def\BlMedianError{\SI{4.4}{\gauss}}

\def\BlChiSqr{2.43}

\def\BlNullChiSqr{0.92}

\def\BlFAP{$10^{-13}$}

\def\BlProt{\SI{46.3 \pm 2.2}{\day}}

\def\BlAmplitude{$4.9^{+1.1}_{-0.9}$\,G}

\def\BlEvolutionTime{$112^{+66}_{-41}$\,d}

\def\MAa{$0.4189^{+0.0073}_{-0.0072}\,M_\odot$}
\def\MAanum{$0.4189^{+0.0073}_{-0.0072}$}

\def\MAb{$0.6180^{+0.0046}_{-0.0045}\,M_\odot$}
\def\MAbnum{$0.6180^{+0.0046}_{-0.0045}$}

\def\Mtot{$1.037\pm0.010\,M_\odot$}

\def\periodYears{$\sim$\SI{9.83}{\year}}
\def\periodYearsFull{$9.829^{+0.033}_{-0.035}$\,yr}

\def\posini{$0.5771 \pm 0.0005$}

\def\periastron{$5.658 \pm 0.001$\,rad}

\def\orbinc{$81.2\pm1.6{}^{\circ}$}
\def\orbincnum{$81.2\pm1.6$}

\def\ascnode{$4.477 \pm 0.009$\,rad}

\def\ascnodeDeg{$193.37\pm0.52$}

\def\infConj{$59\,475.87\pm0.68$}
\def\infConjJD{$2\,459\,475.87\pm0.68$}

\def\vsys{\SI{3313(2)}{\metre\per\second}}

\def\smain{\SI{4.644 \pm 0.017}{\au}}
\def\smainnum{$4.644\pm0.017$}

\def\smaA{\SI{2.760 \pm 0.005}{\au}}

\def\smaB{\SI{1.881 \pm 0.004}{\au}}

\def\ioangle{$81.5^{\circ}$}
\def\ioanglenum{$81.5$}

\def\ioAngleRate{$\sim$0.34\,deg\,yr$^{-1}$}

\def\outerSeparation{\SI{14.64}{\arcsecond}}
\def\outerSeparationnum{$14.64$}

\def\outerPA{$111.9^{\circ}$}
\def\outerPAnum{$111.9$}

\def\outerProjSep{\SI{123.9}{\au}}
\def\outerProjSepnum{$123.9$}

\def\outerPeriod{$>\SI{1000}{\year}$}
\def\outerPeriodnum{$>1000$}

\def\AbTeffH{\SI{5210(320)}{\kelvin}}
\def\AbTeffHnum{$5210\pm320$}

\def\AbTeffHe{$5190\pm340$}

\def\AbLogGH{$8.05 \pm 0.03$}

\def\AbLogGHe{$8.07 \pm 0.03$}

\def\AbProgMassH{$1.44^{+0.38}_{-0.27}$}

\def\AbProgMassHe{$1.43^{+0.38}_{-0.26}$}

\def\AbCoolAgeH{$5.3^{+1.6}_{-1.4}$}
\def\AbCoolAgeHnum{$5.3^{+1.6}_{-1.4}$}

\def\AbCoolAgeHe{$5.52^{+0.91}_{-1.05}$}
\def\AbCoolAgeHenum{$5.52^{+0.91}_{-1.05}$}

\def\AbTotalAgeH{$9.0^{+3.1}_{-2.3}$}

\def\AbTotalAgeHe{$9.0^{+3.1}_{-1.9}$}

\def\AbAltTotalAge{$\sim$\SI{7.8}{\giga\year}}

\def\BTeff{$5075\pm7$}

\def\BLogG{$7.973 \pm 0.004$}

\def\BMass{$0.564\pm0.003$}

\def\BProgMass{$1.22^{+0.21}_{-0.11}$}

\def\BCoolAge{$5.076^{+0.061}_{-0.057}$}

\def\BCoolAgeIntro{\SI{5.08}{\giga\year}}

\def\BTotalAge{$10.8^{+2.2}_{-2.3}$}

\def\IRexcessFlux{$\sim$\SI{3}{\milli\jansky}}

\def\IRdiskTemp{\SI{1200}{\kelvin}}

\def\IRdiskRadius{$\sim$\SI{9}{\earthradius}}

\def\IRdiskWidth{$\sim$\SI{5}{\earthradius}}

\def\HbandResidual{$\sim 1.5\sigma$}

\def\KbandResidual{$\sim 3\sigma$}

\def\AaTeffPhot{\SI{3365(60)}{\kelvin}}

\def\AaTeffSpec{\SI{3510(60)}{\kelvin}}

\def\AaRadius{\SI{0.442 \pm 0.013}{\solarradius}}

\def\AaLogG{$4.770 \pm 0.027$}

\def\sunLogG{$4.4380 \pm 0.0002$}

\def\AaModelLogG{4.75}

\def\AaTeffGrid{3460, 3520, 3580\,K}

\def\AaTeffGaidos{\SI{3363(42)}{\kelvin}}

\def\AaTeffFouque{\SI{3262(19)}{\kelvin}}

\def\AaTeffHejazi{\SI{3350(85)}{\kelvin}}

\def\AaMHFouque{$+0.21 \pm 0.10$}

\def\AaMHHejazi{$-0.25 \pm 0.22$}

\def\AaAlphaHejazi{$+0.12 \pm 0.10$}

\def\AaMHspirou{$+0.17 \pm 0.12$}

\def\AaMHphot{$+0.11 \pm 0.15$}

\def\AaMHphotG{$+0.06 \pm 0.28$}
\def\AaMHphotJ{$+0.12 \pm 0.31$}
\def\AaMHphotH{$+0.14 \pm 0.29$}
\def\AaMHphotK{$+0.11 \pm 0.28$}

\def\AaProtGP{$46.84^{+0.56}_{-0.61}$\,d}

\def\AaGyroAge{\SI{3.6 \pm 1.0}{\giga\year}}

\def\MsixSevenAge{\SI{4}{\giga\year}}

\def\contrastFfourSixSix{$>4.5$\,mag}

\def\sepDecF{\SI{564.6 \pm 3.8}{\mas}}

\def\paDecF{$188.26 \pm 0.62^{\circ}$}

\def\contrastAugF{$6.34 \pm 0.38$\,mag}

\def\sepAugF{\SI{626.1 \pm 2.7}{\mas}}

\def\paAugF{$191.68 \pm 0.41^{\circ}$}

\def\contrastSepFlowSixTwo{$4.63 \pm 0.38$\,mag}

\def\sepSepFlowSixTwo{\SI{639.5 \pm 2.8}{\mas}}

\def\paSepFlowSixTwo{$192.15 \pm 0.41^{\circ}$}

\def\contrastSepFhighEightThree{$6.18 \pm 0.38$\,mag}

\def\sepSepFhighEightThree{\SI{631.1 \pm 2.7}{\mas}}

\def\paSepFhighEightThree{$191.68 \pm 0.41^{\circ}$}

\def\alopekePhotErr{\SI{0.38}{\mag}}

\def\sepChangeTenMonths{$\sim 0.06^{\prime\prime}$}

\def\paChangeTenMonths{$\sim 3^{\circ}$}

\def\nGnirsPerBand{15}

\def\gnirsJexp{$150 \times \SI{0.4}{\second}$}

\def\gnirsHKexp{$300 \times \SI{0.2}{\second}$}

\def\gnirsDither{$3\times 5$ grid of $1.5^{\prime\prime}$ dithered positions}

\def\gnirsZPerr{\SIrange{0.05}{0.10}{\mag}}

\def\AaFilterBluebandpass{540--584}

\def\AaFilterSevenOneSixbandpass{690--742}
\def\AaFilterRedbandpass{812--852}

\def\BBPbandpass{328 - 671}

\def\BRPbandpass{627 - 911}

\def\Bgmag{$14.922\pm0.001$} 
\def\Bgbandpass{420 - 534}

\def\Brmag{$14.367\pm0.001$} 
\def\Brbandpass{556 - 690}

\def\Bimag{$14.16\pm0.01$}
\def\Bibandpass{686 - 840}

\def\Bzmag{$14.089\pm0.004$} 
\def\Bzbandpass{820 - 1006}

\def\Bymag{$13.964\pm0.004$} 
\def\Bybandpass{904 - 1156}

\def\BJmag{$12.972\pm0.065$}

\def\BHmag{$12.803\pm0.060$}

\def\BKmag{$12.654\pm0.058$}

\def\AaJmag{$13.26\pm0.15$}
\def\AaJbandpass{1167 - 1329}

\def\AaHmag{$12.81\pm0.15$}
\def\AaHbandpass{1508 - 1758}

\def\AaKmag{$12.40\pm0.15$}
\def\AaKbandpass{2056 - 2346}

\def\cyclePeriodLTF{$432.8^{+8.1}_{-8.2}$\,d}

\def\cyclePeriodGP{$437^{+15}_{-16}$\,d}

\def\cycleAmpRV{$6.7\pm2.4$\,m\,s$^{-1}$}

\def\cycleAmpFWHM{$6.6^{+5.5}_{-4.5}$\,m\,s$^{-1}$}

\def\cycleAmpDtemp{$1.26^{+0.27}_{-0.26}$\,K}

\def\cyclePhaseRV{$0.16^{+0.81}_{-0.12}$}

\def\cyclePhaseFWHM{$0.46^{+0.13}_{-0.18}$}

\def\cyclePhaseDtemp{$0.410^{+0.070}_{-0.077}$}

\def\cycleAmpFWHMgp{$5.9^{+4.7}_{-4.1}$\,m\,s$^{-1}$}

\def\cycleAmpDtempGP{$0.56^{+0.30}_{-0.33}$\,K}

\def\cyclePhaseFWHMgp{$0.68^{+0.18}_{-0.27}$}

\def\cyclePhaseDtempGP{$0.39^{+0.16}_{-0.15}$}

\def\rvKepPeriod{$3590^{+12}_{-13}$\,d}

\def\rvKepAmp{\SI{8441.4 \pm 4.7}{\metre\per\second}}

\def\rvKepEcc{$0.1898^{+0.0016}_{-0.0015}$}

\def\lnZmep{$-608.4$}
\def\lnZespTwo{$-609.8$}
\def\lnZespThree{$-613.3$}
\def\lnZespFour{$-611.5$}

\def\GPtau{$210^{+150}_{-80}$\,d}

\def\GPeta{$0.33^{+0.10}_{-0.075}$}

\def\GPampFWHM{$28.0^{+7.7}_{-4.8}$\,m\,s$^{-1}$}

\def\GPampDtemp{$1.94^{+0.50}_{-0.30}$\,K}

\def\GPjitterFWHM{$17.3^{+1.7}_{-1.4}$\,m\,s$^{-1}$}

\def\GPjitterDtemp{$0.445^{+0.075}_{-0.066}$\,K}

\def\fwhmDtempSlope{$14.4 \pm 1.4$\,m\,s$^{-1}$\,K$^{-1}$}

\def\priorTYClnA{$\mathcal{LU}(0.056, 560)$}
\def\priorTYClnl{$\mathcal{LU}(10, 1000)$}
\def\priorTYClngamma{$\mathcal{LU}(0.13, 8)$}
\def\priorTYClnP{$\mathcal{LU}(35, 55)$}
\def\priorTYClns{$\mathcal{LU}(0.056, 560)$}
\def\priorTYCslope{$\mathcal{U}(-0.032, 0.032)$}
\def\priorTYCintercept{$\mathcal{U}(-56, 56)$}

\def\postTYClnA{$1.62^{+0.15}_{-0.19}$}
\def\postTYClnl{$4.34 \pm 0.25$}
\def\postTYClngamma{$0.7 \pm 0.4$}
\def\postTYClnP{$3.837^{+0.024}_{-0.031}$}
\def\postTYClns{$-0.15^{+0.28}_{-0.22}$}
\def\postTYCslope{$0.0074 \pm 0.0024$}
\def\postTYCintercept{$0.1 \pm 1.4$}

\def\postLINTYClnA{$5.1^{+0.7}_{-1.0}$}
\def\postLINTYClnl{$76^{+18}_{-21}$}
\def\postLINTYClngamma{$2.0^{+0.7}_{-1.0}$}
\def\postLINTYClnP{$46.4^{+1.1}_{-1.5}$}
\def\postLINTYClns{$0.86 \pm 0.21$}

\def\mcmcNwalkers{100}

\def\mcmcNsteps{10000}

\def\mcmcNburnin{2000}

\def\nestedLivePerParam{40}

\def\nestedSliceSteps{2}

\def\nDetectionBins{100}

\def\detectionPeriodLow{\SI{1}{\day}}

\def\detectionPeriodHigh{\SI{1000}{\day}}

\def\residPeriodRot{\SIrange{42}{45}{\day}}

\def\residPeriodRV{\SI{163}{\day}}

\def\residPeriodFWHM{$\sim$\SI{56}{\day}}

\def\residPeriodAlias{\SI{223}{\day}}

\def\stabilityCtr{0.013}

\def\stabilityMassRatio{$\sim 0.8$}

\def\stabilityRho{$\sim 17$}

\def\stabilityEccMax{$\lesssim 0.80$}

\def\KLtimescale{0.07 and \SI{0.05}{\mega\year}}

\def\preMassLossTotal{\SI{1.86}{\solarmass}}

\def\postMassLossTotal{\SI{1.04}{\solarmass}}

\def\AbProgZAMS{$\sim$\SI{1.4}{\solarmass}}

\def\massShed{$\sim$\SI{0.8}{\solarmass}}

\def\preMassLossSMA{$\sim$\SI{2.6}{\au}}

\def\AGBtipMbol{$-4.52$}

\def\AGBtipRadius{\SI{1.2}{\au}}

\def\AGBtipTeff{\SI{3000}{\kelvin}}

\def\massSiriusB{\SI{1.018 \pm 0.011}{\solarmass}}

\def\massProcyonB{\SI{0.592 \pm 0.006}{\solarmass}}

\def\masseridB{\SI{0.573 \pm 0.018}{\solarmass}}

\def\MdwarfTypicalErr{$5$--$10\%$}

\def\GtwoZeroThreeDistance{\SI{7.3}{\parsec}}

\def\GtwoZeroThreePeriod{\SI{14.7}{\day}}

\def\GtwoZeroThreeSMA{\SI{15}{\solarradius}}

\def\GtwoZeroThreeSMAAU{\SI{0.05}{\au}}

\def\GtwoZeroThreeWDmass{$\gtrsim$\SI{0.5}{\solarmass}}

\def\GtwoZeroThreeMH{$\simeq +0.14$}

\def\encounterRate{$\sim 3\times 10^{11}$\,yr}

\def\outerSkyAccel{$\sim 0.3$\,mas\,yr$^{-2}$}

\def\POSSbaseline{$\sim$\SI{75}{\year}}

\def\POSSpaChange{$\sim 25^{\circ}$}

\def\cyclePeriodLTFnum{$432.8^{+8.1}_{-8.2}$}

\def\cycleAmpRVnum{$6.7\pm2.4$}

\def\cycleAmpFWHMnum{$6.6^{+5.5}_{-4.5}$}

\def\cycleAmpDtempnum{$1.26^{+0.27}_{-0.26}$}

\def\rvKepPeriodnum{$3590^{+12}_{-13}$}

\def\rvKepAmpnum{$8441.4\pm4.7$}

\def\ltfSysVel{$(-3.083^{+0.011}_{-0.012})\times10^{3}$}

\def\ltfFWHMlinSlope{$(3.86^{+0.49}_{-0.52})\times10^{-2}$}

\def\ltfFWHMoffset{$(6.13^{+0.82}_{-0.86})\times10^{1}$}

\def\ltfDtempLinSlope{$(2.80^{+0.31}_{-0.29})\times10^{-3}$}

\def\ltfDtempOffset{$4.33\pm 0.50$}

\def\ltfRVjitter{$7.61^{+0.67}_{-0.58}$}

\def\ltfFWHMjitter{$28.1^{+2.2}_{-2.0}$}

\def\ltfDtempJitter{$1.57^{+0.13}_{-0.11}$}

\def\rvKepPhase{$0.4208^{+0.0014}_{-0.0013}$}

\def\rvKepOmega{$5.6615^{+0.0083}_{-0.0088}$}

\def\cyclePeriodGPnum{$437^{+15}_{-16}$}

\def\cycleAmpFWHMgpnum{$5.9^{+4.7}_{-4.1}$}

\def\cycleAmpDtempGPnum{$0.56^{+0.30}_{-0.33}$}

\def\gpFWHMlinSlope{$(3.7\pm1.1)\times 10^{-2}$}

\def\gpFWHMoffset{$(6.3\pm1.7)\times 10^{1}$}

\def\gpDtempLinSlope{$(2.63^{+0.76}_{-0.71})\times 10^{-3}$}

\def\gpDtempOffset{$4.4^{+1.2}_{-1.1}$}

\def\GPjitterFWHMnum{$17.3^{+1.7}_{-1.4}$}

\def\GPjitterDtempnum{$0.445^{+0.075}_{-0.066}$}

\def\GPtaunum{$210^{+150}_{-80}$}

\def\AaProtGPnum{$46.84^{+0.56}_{-0.61}$}

\def\GPampFWHMnum{$28.0^{+7.7}_{-4.8}$}

\def\GPampDtempnum{$1.94^{+0.50}_{-0.30}$}

\def\IRexcessNmatched{701}
\def\IRexcessWoneOffset{0.17}
\def\IRexcessWtwoOffset{0.16}
\def\IRexcessPoolScatter{0.22--0.24}
\def\IRexcessTypicalOffset{0.13}
\def\IRexcessWthreeOffset{0.04}
\def\IRexcessAKARIoffset{0.13}
\def\IRexcessIRASoffset{0.36}
\def\IRexcessIRASratio{39\%}

\def\SteinMass{$0.675\pm0.051\,M_\odot$}
\def\SteinMassNum{$0.675\pm0.051$}
\def\SteinRadius{$1.14\pm0.04$}
\def\SteinTeff{$7120\pm180$}
\def\SteinCoolAge{$\sim$2.0}
\def\LawdMass{$0.56\pm0.08\,M_\odot$}
\def\LawdMassNum{$0.56\pm0.08$}
\def\LawdRadius{$1.27\pm0.03$}
\def\LawdTeff{$7837\pm83$}
\def\LawdCoolAge{$1.15\pm0.04$}
\def\SiriusCoolAge{$\sim$0.13}
\def\ProcyonCoolAge{$\sim$1.2}
\def\EridBCoolAge{$\sim$0.11}

\abstract
{We report the serendipitous discovery of a white dwarf companion to the M4\,V dwarf \tyc\ (IRAS~21500+5903) during a radial-velocity search for exoplanets with SPIRou. The system is located at \sysDistance\ and forms a hierarchical triple with a more distant white dwarf, which has no known analogue within \SI{30}{\parsec}.}
{We characterise the orbital architecture and component masses through a joint Keplerian fit of radial velocities and astrometry, derive the physical parameters of the M dwarf and both white dwarfs, and discuss the implications for common-envelope evolution and hierarchical triple dynamics.}
{We obtained radial velocity measurements with SPIRou spread over \durationRv, revealing a large-amplitude signal ($\sim$\rvPeakToPeak\ peak-to-peak) that is inconsistent with planetary companions. Follow-up speckle imaging with \AlopekeH\ at Gemini North and near-infrared (NIR) imaging with GNIRS confirmed a white dwarf companion at sub-arcsecond separation. We performed a joint Keplerian fit combining radial velocities and relative astrometry, yielding high-precision orbital parameters and component masses.}
{The system comprises an inner binary with a semi-major axis of \smain\ ($P \approx 9.83$\,yr), consisting of the primary M4\,V ($M =$ \MAa) and a white dwarf companion ($M =$ \MAb, $T_{\rm eff} \sim$ \AbTeffH, cooling age \SIrange{5.3}{5.5}{\giga\year}). This inner pair lies at a projected separation of \outerProjSep\ ($P$~\outerPeriod) from a previously known white dwarf. The dynamically determined mass of the inner white dwarf is one of only a handful of such measurements for white dwarfs within \SI{20}{\parsec}, and the most precise, at the sub-percent (0.7\%) level.}
{This system is the closest known triple containing multiple white dwarfs and offers direct dynamical constraints on common-envelope evolution and hierarchical triple stability. Furthermore, a tentative near-infrared (IR) excess towards the inner white dwarf further hints at circumstellar dust.}

   \keywords{stars: individual: IRAS 21500+5903 -- stars: low-mass -- binaries: close -- white dwarfs -- techniques: radial velocities -- techniques: high angular resolution}

   \maketitle
\nolinenumbers

\section{Introduction}

Serendipity has long played a key role in astronomy, offering insights into new phenomena, from Galileo's unexpected observation of Jupiter's four largest moons with his newly constructed refractor telescope in 1610 \citep{galilei_sidereus_1610} to Herschel's discovery of Uranus, which he first reported as a comet when examining faint stars as part of his observations on stellar parallax \citep{herschel_account_1781}.  More recent examples include the discovery of Boyajian's Star (KIC 8462852) through a citizen science examination of \textit{Kepler} photometry \citep{boyajian_planet_2016}  and the detection of the interstellar object 1I/'Oumuamua \citep{meech_brief_2017}.

Serendipity is an accepted mode of progress in observational astronomy \citep{rond_serendipity_2010}. Surveys of large samples, whatever their primary goal, frequently yield unexpected discoveries \citep{norris_discovering_2017} and radial-velocity (RV) surveys for exoplanets are no exception. The discovery of the first exoplanet orbiting a main sequence star \citep{mayor_jupiter-mass_1995} was unexpected: formation models placed giant planets beyond $\sim$5\,au, on orbits of a decade or longer, but the survey instead revealed a hot Jupiter on a \SI{4.2}{\day} orbit, a configuration that these models had not anticipated. This discovery reshaped our understanding of planetary system architectures and demonstrated the power of high-precision RV monitoring to reveal unexpected configurations.

We report the serendipitous discovery of a white dwarf (WD) companion to an M dwarf in the solar vicinity ($d\sim8$\,pc), uncovered during a blind RV planet search. The system is unique among nearby examples: it is a hierarchical triple made of one M dwarf and two WDs, in which the M dwarf may have survived common-envelope evolution and it has no known analogue within \SI{30}{\parsec}. The discovery via RV is discussed in Sect.~\ref{discovery}, the activity of the M dwarf in Sect.~\ref{activity}, direct imaging follow-up in Sect.~\ref{imaging}, Keplerian orbital fitting in Sect.~\ref{orbital_fit}, and the properties of the M dwarf and of the inner WD are presented in Sects.~\ref{properties_m_dwarf} and~\ref{properties_white_dwarf}, respectively. We discuss the secular evolution of the system in Sect.~\ref{conclusion} and present our conclusions in Sect.~\ref{sec:conclusions}.

\section{Serendipitous discovery of the TYC~3980-1081-1 companion\label{discovery}}

SPIRou is a near-infrared (NIR) spectropolarimeter \citep{donati_spirou_2018,donati_spirou_2020} stabilised for precision RV measurements. SPIRou operates over the $YJHK$ bandpasses, with a spectral resolution of $\lambda/\Delta\lambda \sim 70\,000$ \citep{donati_spirou_2020}, and its RV performance is at the \SI{2}{\metre\per\second} level \citep{artigau_line-by-line_2022}.

To maximise the scientific return of SPIRou, multi-year observing efforts have been awarded time as Large Programmes (LPs). The first SPIRou LP, the SPIRou Legacy Survey \citep[SLS;][]{donati_spirou_2018,donati_spirou_2020}\footnote{All programme, run, and observation IDs referenced in this paper are listed in the Acknowledgements.}, was organised into three work packages (WPs): WP1 targeted nearby M dwarfs to characterise the architecture of their planetary systems, WP2 focussed on follow-up observations of known transiting planets, and WP3 investigated the physics of pre-main sequence stars. Within WP1 and WP2, several planetary systems have already been published \citep[e.g.][]{cadieux_toi-1452_2022,martioli_toi-1759_2022,cortes-zuleta_gi_2025,moutou_characterising_2024,ould-elhkim_spirou_2025,ould-elhkim_spirou_2026,Charpentier_gj4274_2026,Carmona_gl410_2025}, the activity and magnetic fields of individual targets have been characterised in detail \citep[e.g.][]{cortes-zuleta_optical_2023,donati_mdwarfs_2023}. Additional systems are under continued monitoring to confirm tentative Keplerian signals.

Since 2019, one of the systems monitored for low-mass planets as part of SLS WP1 exhibited a highly unusual RV signature: IRAS~21500+5903 (TYC~3980-1081-1). The mid-M dwarf showed a near-constant acceleration of approximately \SI{5}{\metre\per\second\per\day} ($\sim$\SI{1800}{\metre\per\second\per\year}), a level inconsistent with a planetary companion: on a few-year orbit around a \SI{0.42}{\solarmass} star, even a Jupiter-mass planet would induce an acceleration of at most $\sim$40--130\,\si{\metre\per\second\per\year}. Before an accurate spectral type (M3.9\,V) was established by \citet{fouque_spirou_2018}, the mid-infrared (MIR) detection that gave the star its designation in the Infrared Astronomical Satellite (IRAS) catalogue (discussed further in Sect.~\ref{sec:excess}), the absence of a pre-\textit{Gaia} parallax and an uncertain optical classification led the CARMENES team to suspect it was a background giant, rather than a nearby M dwarf; as a result, this otherwise unremarkable M dwarf (in 2024, the only M dwarf within 9\,pc in the CARMENES input catalogue still lacking both RV monitoring and known close companions; \citealt{Caballero2024csss}) had not been included in other RV planet-search programmes when SLS began monitoring it. The mislabelling plausibly explains why the system had not been identified sooner, despite having been detectable even with early-generation RV spectrographs.

The initial interpretation suggested the presence of a previously unidentified close-in brown dwarf ($s<1$\,au) on an orbit of approximately \SI{1}{\year}. The corresponding projected separation of about \SI{100}{\mas} would have explained why the companion had not been detected in previous imaging surveys. The system was monitored over the remaining three years of SLS and was subsequently included in the follow-up LP, SPIRou Legacy Survey $-$ Consolidation \& Enhancement (SPICE), which extended the SLS time baseline. By mid-2023, the RV signal began to show curvature, enabling meaningful constraints on the companion's orbit. The data indicated a decade-long orbital period and a companion mass well in excess of the upper limit expected for a substellar object. The inferred mass, $\gtrsim\SI{0.6}{\solarmass}$, exceeds that of the $\sim\SI{0.42}{\solarmass}$ M-dwarf primary. A single main sequence companion could be ruled out, as it would have to be a K dwarf or earlier, and therefore much brighter than the M dwarf itself in both the optical and infrared (IR) regimes. A binary mid-M companion could have explained the properties of the system but this would have required both components to be fast rotators, for the pair not to be resolved as a double-lined (SB2) spectroscopic binary.

We concluded that the companion was most likely, in decreasing order of probability and increasing orbital inclination: a WD, a neutron star, or a black hole. This motivated a detailed investigation to detect the companion directly or, at a minimum, to place meaningful upper limits on its flux.

At \mbox{$\sim\SI{8}{\parsec}$}, the system was already known to host a DAH (hydrogen-dominated and magnetic) WD companion, WD~2150+591 \citep{caron_spectrophotometric_2023}. The wide pair was first reported as a visual double by Knapp (discovery code KPP~4252 in the Washington Double Star Catalog, where it is listed as WDS~21516+5918; \citealt{mason_2001_2001}). The pair is one of eight systems within \SI{10}{\parsec} known to combine a WD with a non-degenerate star (seven with a resolved WD; \citealt{gonzalezpayo_multiple_2026}), and their proper motions and parallaxes point to a physical association, though neither matches precisely: the M~dwarf's own \textit{Gaia} Data Release 2 (DR2) proper motion and parallax are each measurably offset from the outer WD's \textit{Gaia} Data Release 3 (DR3) values (the M~dwarf's DR2 parallax, $\varpi_{Aa}$, is 4.15\% higher than $\varpi_B$, and its declination proper motion differs in sign), and its DR2 astrometric solution is independently flagged as of poor quality (\texttt{astrometric\_chi2\_al} = \AaAstrometricChiSq), a discrepancy we return to quantitatively in Sect.~\ref{sec:keplerian_motion}; part of the mismatch may also reflect the relative orbital motion of the M dwarf and the distant WD around their common barycentre (comparable in mass, neither can be treated as stationary), though this is expected to be small given the outer period, of the order of $10^3$\,yr. At a separation of $\sim$\SI{120}{\au}, even a maximally favourable edge-on orbit would induce an acceleration of only $\sim$\SI{1}{\metre\per\second\per\year} on its primary, far too small to explain the RV signal that revealed the inner companion. Despite these small kinematic mismatches, the two objects are almost certainly physically linked. Setting these mismatches aside, the sky-plane proximity of the pair is itself difficult to explain by chance: of the 461 individual objects catalogued within 10\,pc of the Sun \citep{kirkpatrick_initial_2024}, the probability that any given one would have an unrelated 10-pc object fall, purely by chance projection, within the observed \outerSeparation\ separation is $\sim6\times10^{-7}$; allowing for the fact that any of the 461 objects could in principle have been the one found close to any other only raises this to $\sim3\times10^{-4}$.

Crucially, the outer WD provides a precise \textit{Gaia} DR3 distance to the system \citep{gaia_collaboration_dr3_2023}, whereas the M dwarf's own \textit{Gaia} DR2 parallax (\AaParallaxDRtwo) is unreliable for the reasons given above, pending a full orbit solution of the inner binary (likely with the upcoming \textit{Gaia} DR4). We therefore adopted the outer WD's parallax throughout; Sect.~\ref{sec:keplerian_motion} quantifies how sensitive the derived system properties are to this choice.

TYC~3980-1081-1, the outer WD (WD~2150+591) and this newly identified inner massive companion constitute a hierarchical triple system, hereafter named \welo\ (the origin and meaning of this Hawaiian name are discussed in Appendix~\ref{sec:name}). We designated the M dwarf as component~Aa (\tyc), the inner massive companion as component~Ab (\innerwd), and the outer WD as component~B (\outerwd). Throughout this paper we use the term `primary' for \tyc\ in a liberal sense: the M dwarf is by far the brightest object in the system at the wavelengths considered; consequently, \innerwd\ and \outerwd\ are both designated as `companions', even though they are more massive than \tyc.

\begin{figure}[!t]
    \centering
    \includegraphics[width=\linewidth]{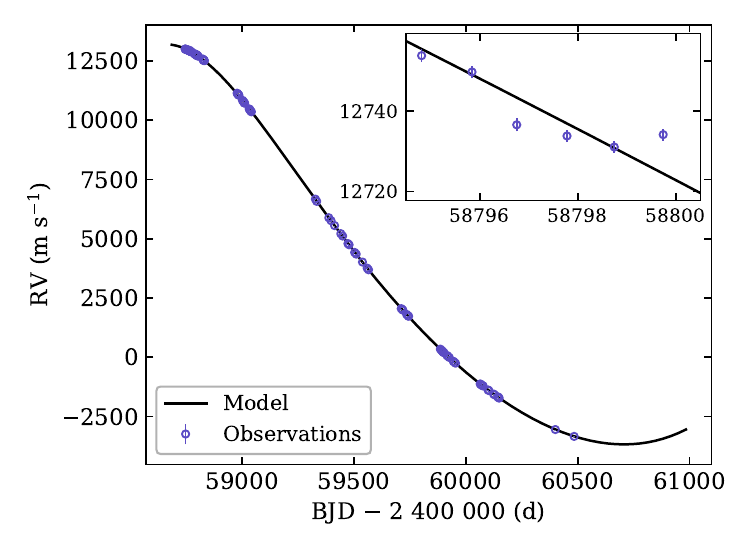}
\caption{RV curve of \tyc\ from SPIRou observations spanning \durationRv, with the best-fit Keplerian model overplotted. Inset: zoom on a week-long observing sequence showing the measurement precision and sampling cadence; each point is the average of the four integrations of one observing date, so its error bar, \spirouMedianNightlyRVerr\ in the median, is half the \spirouMedianRVerr\ uncertainty of an individual integration quoted in Sect.~\ref{discovery}.}
    \label{fig:rv_postcard}
\end{figure}

SPIRou observations of  \tyc\ consist of \nRVframes\ validated integrations obtained in polarimetric sequences of four frames (one frame did not pass quality control, so the total is not a multiple of four), collected over \nRVepochs\ epochs spanning a total of \durationRv.
The data reduction was performed with the \texttt{APERO} pipeline, version \aperoVersion\ \citep{cook_apero_2022}. \texttt{APERO} handles all aspects of precision RV extraction: preprocessing of raw frames, localisation and calibration of order traces, optimal extraction, wavelength calibration, and telluric correction. RV measurements were performed within the line-by-line (LBL) framework \citep{artigau_line-by-line_2022}. The \texttt{APERO} and LBL frameworks have demonstrated \si{\metre\per\second}-level performance on a number of SPIRou targets \citep[e.g.][]{cadieux_toi-1452_2022,martioli_toi-1759_2022}. The observations have a median $H$-band signal-to-noise ratio (S/N) of \spirouMedianSNR, leading to a per-spectrum RV uncertainty of \spirouMedianRVerr. While the optimisation of RV precision is central to most SPIRou exoplanet applications, its impact here is minimal, as the RV signal is nearly four orders of magnitude larger than the median error. Figure~\ref{fig:rv_postcard} shows the RV measurements of \tyc, with a peak-to-peak change of about \rvPeakToPeak.

\section{Analysis and results\label{sec:analysis_results}}

\subsection{Activity and large-scale magnetic field of \tyc\label{activity}\label{magfield}}

All SPIRou observations of \tyc\ were carried out in polarimetric mode and were also processed with {\tt Libre ESpRIT}, the nominal reduction pipeline of ESPaDOnS at the Canada-France-Hawai\okina i Telescope (CFHT) \citep{donati_spectropolarimetric_1997}, adapted for SPIRou \citep{donati_spirou_2020} and optimised for spectropolarimetry \citep{donati_mdwarfs_2023}.  Each visit of \tyc\ consisted of sequences of four subexposures aimed at measuring circularly polarised (Stokes $V$) and unpolarised (Stokes $I$) profiles of spectral lines, enabling probing of the large-scale magnetic field at the stellar surface through the Zeeman effect \cite[][]{donati_spectropolarimetric_1997}.  From each sequence, we also derived a null polarisation check (called $N$) that should be equal to zero down to photon noise level if the polarimeter behaved correctly, thus allowing us to diagnose potential sources of spurious polarisation from either the instrument or data reduction.  We collected \nStokesVtotal\ Stokes $V$ sequences of \tyc, six of which were discarded due to poor weather conditions.

From the \nStokesVkept\ remaining Stokes $I$, $V$, and $N$ spectra, we computed the least-squares deconvolved (LSD) profiles \citep{donati_spectropolarimetric_1997} using a line mask adequate to an M3 dwarf, then derived the longitudinal field \Bl\ from the first moment of the Stokes $V$ LSD profiles normalised by the equivalent widths of the Stokes $I$ LSD profiles \citep[integrating over an interval of $\pm$13~\kms, as in][]{donati_mdwarfs_2023}.  The inferred \Bl\ values vary from \BlRange\ (median \BlMedian), with a median error bar of \BlMedianError\ (Table~\ref{tab:bl}).  \Bl\ values derived from the $N$ (instead of $V$) LSD profiles in the same way are all consistent with zero, confirming that the detected polarisation is not spurious.   The large-scale field of \tyc\ is detected at high significance: with the reduced chi-square \chisqr\ of the \Bl\ curve reaching \BlChiSqr\ for the \nStokesVkept\ measurements, corresponding to a false-alarm probability of \BlFAP\ that the detection is spurious (the same operation on \Bl\ values derived from $N$ yielding a \chisqr\ of \BlNullChiSqr).

We carried out a Gaussian process (GP) regression fit to the \Bl\ curve using a quasi-periodic covariance function \citep{rajpaul_gaussian_2015}, as done by \citet{donati_mdwarfs_2023} for other M dwarfs observed with the SLS. This fit yielded a significant detection of the rotation period of \tyc, $P_{\rm rot}=$~\BlProt: this value is consistent with the rotation period of \AaProtGP\ we adopted for \tyc. This value was derived independently (as described in Sect.~\ref{sec:rotation_age_discussion} and Appendix~\ref{sec:methods_gp}) from the time series of two spectroscopic activity indicators: the full width at half maximum (FWHM) of the spectral lines and the differential temperature (\textit{dTemp}; \citealt{artigau_measuring_2024}). More specifically, the \Bl\ curve of \tyc\ was found to oscillate more or less symmetrically with respect to the $\Bl=0$ line with an average semi-amplitude of \BlAmplitude, and to evolve on a timescale of \BlEvolutionTime. This near-symmetric oscillation about zero suggests that the large-scale field of \tyc\ is mostly non-axisymmetric with respect to the rotation axis, resembling that of other M3 dwarfs of similar rotation periods \citep[e.g. Gl~687 and Gl~849, with $P_{\rm rot}\simeq57$ and 42\,d;][]{donati_mdwarfs_2023}. Thus, the highly peculiar evolutionary history of \tyc\ does not appear to be manifested in its large-scale magnetic field.

\subsection{Astrometric and photometric follow-ups \label{imaging}}
Considering the unknown inclination of the orbital plane, the RV measurements only provided a lower limit on the mass of the companion. This prompted us to request a Director's Discretionary Time allocation at Gemini North to obtain speckle imaging of \tyc\ and either confirm the WD hypothesis or prompt follow-ups at other wavelengths for more speculative scenarios involving a neutron star or a black hole.

\subsubsection{\AlopekeH\label{sec:alopeke}}

\AlopekeH\ is a dual-channel speckle imager on the Gemini North \SI{8.1}{\metre} telescope \citep{scott_nessi_2018,scott_twin_2021}. \AlopekeH\ performs simultaneous high-resolution imaging in two optical bands over a \SI{6.7}{\arcsec} field of view, achieving the diffraction limit of the telescope ($\sim$0.016--$0.025^{\prime\prime}$ between 500 and 800\,nm) under optimal conditions. Fast frame readout (exposure times of \SI{60}{\milli\second}) allows reconstruction of diffraction-limited images via speckle interferometry, enabling detection of close stellar companions down to separations of a few tens of milliarcseconds. Speckle imaging of this kind is particularly well suited for stellar multiplicity studies and the validation of candidate exoplanets \citep{howell_speckle_2011,scott_nessi_2018}. In the WD scenario, we expected a \numrange{5}{7} magnitude contrast at optical wavelengths and a \SIrange{0.5}{1}{\arcsec} separation (depending on orbital inclination), and a non-detection would have ruled out all reasonable WD scenarios.

\begin{figure}[!t]
    \centering
    \includegraphics[width=\linewidth]{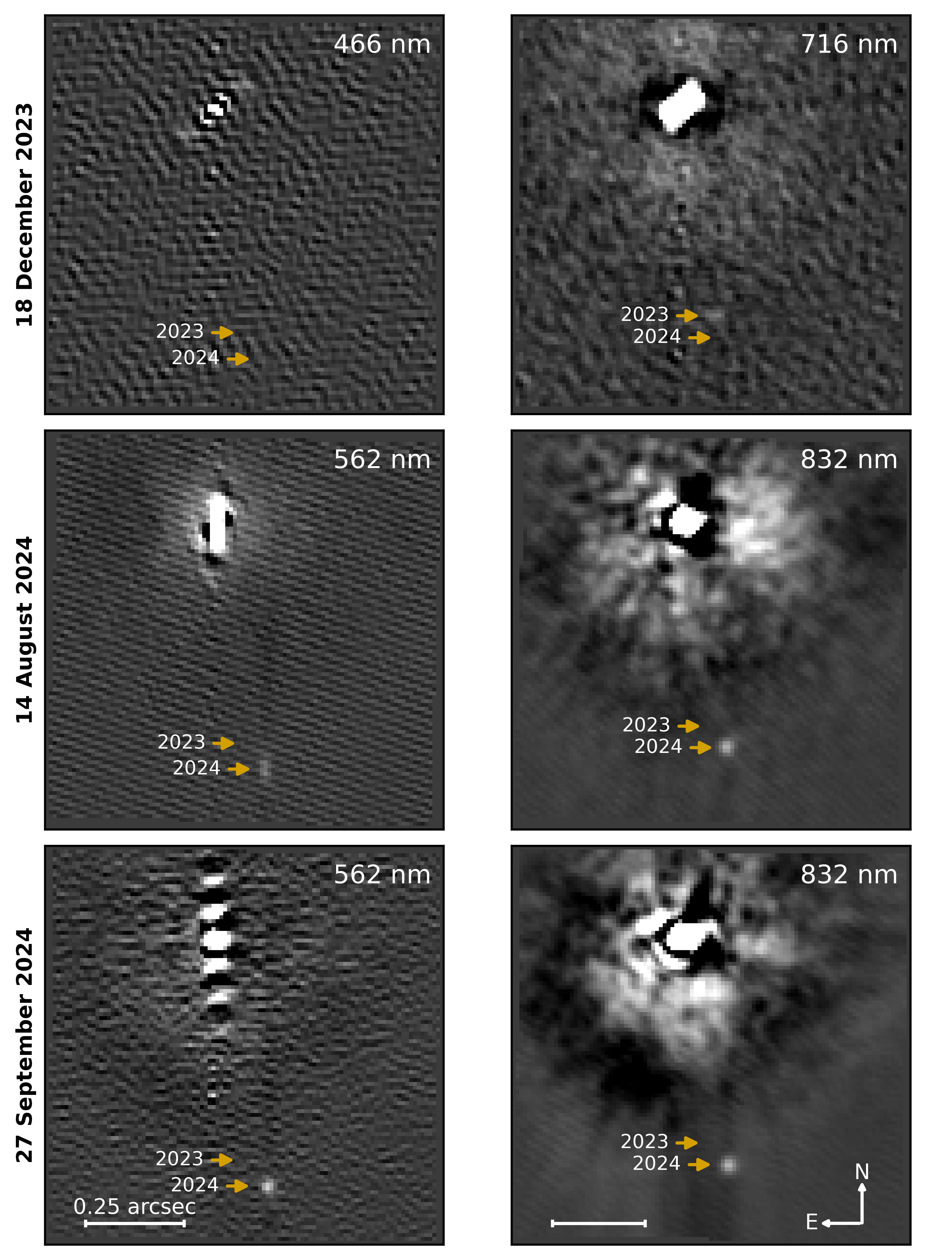}
    \caption{\AlopekeH\ observations of \tyc\
    at three epochs. The yellow arrows mark the position of the WD in December 2023 and in 2024, highlighting the increase in separation; the 2024 arrow points to the mean of the August and September 2024 positions, which agree to within about one pixel. The WD is undetected in the 466\,nm observation obtained in December 2023 and only marginally detected at 562\,nm in August 2024, but it is detected at high significance at 832\,nm in August 2024 and in both bands in September 2024. The two \AlopekeH\ channels have slightly different plate scales, so each column carries its own \SI{0.25}{\arcsecond} scale bar. The scaling is stretched to highlight the WD, but note that the M dwarf is unsaturated in the full dynamic range of the image.}
    \label{fig:alopeke_pbs}
\end{figure}

\AlopekeH\ observations of \tyc\ were obtained on 18 December 2023. The observations were performed with bandpasses centred at \SI{466}{\nano\metre} and \SI{716}{\nano\metre}.
A faint source was detected in the \SI{716}{\nano\metre} data at a contrast of \mbox{$\sim$\SI{6}{\mag}}, prompting follow-up observations. Two additional observations were obtained on 14 August 2024 and 27 September 2024 in bandpasses centred at \SI{562}{\nano\metre} and \SI{832}{\nano\metre}.
These observations are shown in Fig.~\ref{fig:alopeke_pbs} and provide a highly significant detection of orbital motion between epochs.

The data products were reduced by the \AlopekeH\ team, following the standard speckle reduction procedure \citep{horch_dssi_2012,scott_twin_2021}.
Differential astrometry and photometry of \innerwd\ relative to \tyc\ were then performed on the observatory-provided combined frames: we fitted a two-dimensional Gaussian profile to the point spread function (PSF) of \tyc\ and used the derived parameters to fit a Gaussian profile, with the same widths and orientation, to \innerwd. Associated uncertainties were determined through injection and recovery of Gaussians of the same amplitude at random positions in the field. We were unable to retrieve useful constraints from the \SI{466}{\nano\metre} image (source undetected) and from the \SI{562}{\nano\metre} observation of 14 August 2024 (marginal detection only); as the longer-wavelength channels had much more significant detections at all epochs, only these data were used for the astrometric fit (Sect.~\ref{orbital_fit}). Astrometric measurements used the plate scale and orientation provided by the \AlopekeH\ team, whose uncertainties \citep{horch_observations_2019} were propagated through the Gaussian fit and added in quadrature. The resulting contrasts, separations, and position angles are listed in Table~\ref{tab:contrast_separation}.

\begin{figure}[!t]
    \centering
    \includegraphics[width=\linewidth]{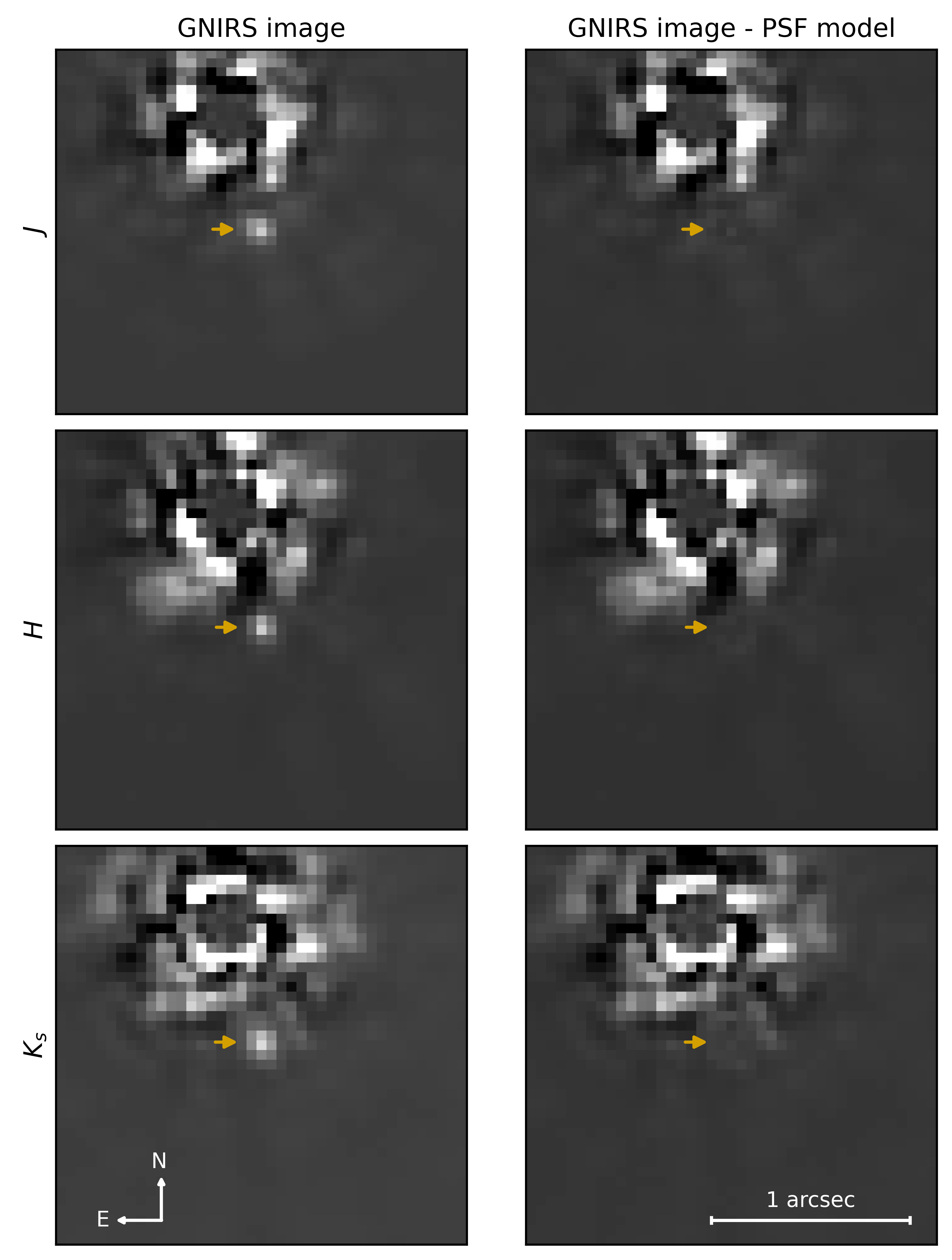}
    \caption{GNIRS NIR imaging of the \tyc\ system in $J$, $H$, and $K_s$ bands, in a region centred on the inner WD companion (yellow arrows), before (left) and after (right) subtraction of a PSF model of the WD. The inner WD companion is detected in all three bands. The structure above the WD is the residual of the strongly saturated PSF of \tyc, whose radial profile was subtracted.}
    \label{fig:gnirs_images}
\end{figure}

Converting contrasts into magnitudes is non-trivial, as \tyc\ does not have published photometry in the non-standard \AlopekeH\ bandpasses. Furthermore, we could not find a spectrophotometrically calibrated archival spectrum of \tyc. We anchored the photometry of \innerwd\ by first retrieving the average M4\,V spectrum from the Sloan Digital Sky Survey (SDSS) spectral library \citep{bochanski_low-mass_2007} and performing spectrophotometric calibration of the spectrum to the \tyc\ \textit{Gaia} magnitude \citep[$G=$ \AaGmag, including an adopted 0.010\,mag calibration floor;][]{gaia_collaboration_dr3_2023}. This allowed us to determine approximate magnitudes, reported in Table~\ref{tab:astrometry_photometry}, which served as an anchor for the conversion of contrasts to magnitudes.

\subsubsection{GNIRS imaging}\label{sec:gnirs}

In addition to its spectroscopic capabilities, the Gemini Near-Infrared Spectrograph (GNIRS; \citealt{elias_performance_2006}) can be used in imaging mode to obtain NIR photometry over a keyhole-shaped field of view.\footnote{\url{https://www.gemini.edu/instrumentation/gnirs/capability}} GNIRS offers imaging at 1.0--2.5\,\si{\micro\metre} (plus a 3.3\,\si{\micro\metre} filter); with the long blue camera, used here, the plate scale is \SI{0.05}{\arcsec} per pixel.

Observations were obtained on 8 September 2024.
Considering the brightness of \tyc, we attempted to minimise saturation and obtained \nGnirsPerBand\ images in each of $J$, $H$, and $K_s$ with \gnirsJexp\ ($J$) or \gnirsHKexp\ ($H$ and $K_s$). Frames were obtained on a \gnirsDither, allowing for the removal of detector artefacts. Images were dark-subtracted, flat-fielded, registered, and median-combined. Figure~\ref{fig:gnirs_images} illustrates a cropped region centred on the WD companion that is detected in all three bands. The core of the PSF of \tyc\ is strongly saturated, so no direct contrast measurement could be made. 

To calibrate the photometry, we identified several field stars within the GNIRS field of view that have United Kingdom Infrared Telescope (UKIRT) Infrared Deep Sky Survey (UKIDSS) DR11 \citep{lawrence_ukirt_2007} archival imagery and published magnitudes; these were used to anchor the photometry of \innerwd. The photometric zero-point was determined from the UKIDSS calibration stars, with typical uncertainties of \gnirsZPerr\ in each band. The final photometric measurements are presented in Table~\ref{tab:astrometry_photometry} (Appendix~\ref{sec:astrometric_table}), which compiles the astrometric and photometric properties of all three components of \welo.

\begin{table}[!t]
    \centering
    \caption{\AlopekeH\ relative astrometry and photometry.}
    \footnotesize
    \setlength{\tabcolsep}{4pt}
    \begin{tabular}{lcccc}
    \toprule
    \toprule
        Date  & Filter & $\Delta m$ & $\rho$ & $\theta$ \\
                  &  & (mag) & (mas) & ($^\circ$) \\

    \midrule
        18 Dec. 2023 & F466 &   $>4.5$\tablefootmark{a}&  \multicolumn{2}{c}{undetected}  \\
        18 Dec. 2023 & F716 & unreliable\tablefootmark{b} & $564.6 \pm 3.8$ & $188.26 \pm 0.62$ \\
        14 Aug. 2024 & F562 &\multicolumn{3}{c}{marginal detection; $<3\sigma$}\\
        14 Aug. 2024 & F832 & $6.34 \pm 0.38$\tablefootmark{c} & $626.1 \pm 2.7$ & $191.68 \pm 0.41$ \\
        27 Sep. 2024 & F562 & $4.63 \pm 0.38$\tablefootmark{c} & $639.5 \pm 2.8$ & $192.15 \pm 0.41$ \\
        27 Sep. 2024 & F832 & $6.18 \pm 0.38$\tablefootmark{c} & $631.1 \pm 2.7$ & $191.68 \pm 0.41$ \\
    \bottomrule
    \end{tabular}

    \tablefoot{\tablefoottext{a}{3$\sigma$ upper limit.} \tablefoottext{b}{The PSF has significant power in its alias at 180$^\circ$.} \tablefoottext{c}{0.38\,mag photometric uncertainty from \citet{clark_pokemon_2024}.}}
    \label{tab:contrast_separation}
\end{table}

\subsection{Keplerian fit combining RV and imaging \label{orbital_fit}}
\label{sec:keplerian_motion}

The orbital parameters of the inner binary were derived through a joint fit detailed in Appendix~\ref{sec:methods_ltf}: a Keplerian orbit was simultaneously adjusted with a long-term activity cycle and polynomial baseline trends fitted to the SPIRou RV, FWHM, and \textit{dTemp} time series (Fig.~\ref{fig:raw_timeseries}; the fit itself is shown in Fig.~\ref{fig:ltf_model_timeseries}). This approach separates the large-amplitude Keplerian signal from stellar-activity contributions. Figure~\ref{fig:rv_postcard} shows the resulting RV time series with the best-fit Keplerian model superimposed.

As shown in Fig.~\ref{fig:rv_postcard} and discussed in Appendix~\ref{sec:methods_gp}, the RV residuals to this fit are dominated by a jitter term of \ltfRVjitter\,\si{\metre\per\second}, well above the median formal RV uncertainty (\spirouMedianRVerr). As detailed in Appendix~\ref{sec:methods_gp}, we were unable to model this residual with a GP stellar-activity kernel, in either a one- or multi-dimensional configuration, because the \innerwd\ orbital signal (peak-to-peak \rvPeakToPeak) is more than three orders of magnitude larger than the residual scatter, which prevents the optimiser from usefully separating the two. The jitter was instead absorbed as excess white noise in the Keplerian fit. Because it is treated as uncorrelated from epoch to epoch, and the $\sim$10-yr orbital period (Table~\ref{tab:mass_m_dwarf}) is sampled over \nRVepochs\ epochs spanning \durationRv, we do not expect this jitter to bias the best-fit period, semi-amplitude, or eccentricity: an unmodelled correlated (e.g. rotation-period-scale) activity signal could, in principle, alias into the orbital solution, but at \AaProtGP, \tyc's rotation period is almost two orders of magnitude shorter than the orbital period (a factor of $\sim$80), making such an alias implausible.

\begin{figure}[!t]
    \includegraphics[width=\linewidth]{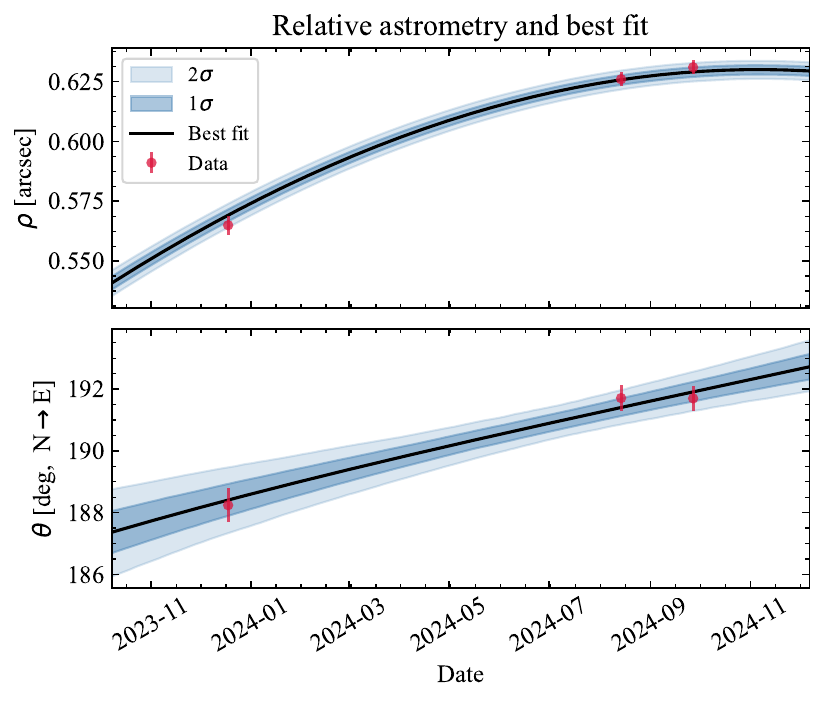}
    \caption{Relative astrometry of the inner binary and best-fit orbital model
             from 20\,000 Monte Carlo samples.
             Top: Angular separation, $\rho$. Bottom: position angle, $\theta$ (north through east).  The dark blue band shows the 1$\sigma$ (68\%) credible interval and the light band the 2$\sigma$ (95\%) interval; red points are the three \AlopekeH\ epochs used in the fit (Table~\ref{tab:contrast_separation}: the 832\,nm measurements, and the 716\,nm one in December 2023) and the black line is the best-fit model.}
    \label{fig:kep_fit}
\end{figure}

We determined the individual masses of the inner binary independently of the
joint orbital fit, using the orbital parameters derived in
Appendix~\ref{sec:methods_ltf}, the \textit{Gaia} DR3 parallax of \outerwd, and
three epochs of \AlopekeH\ relative astrometry.
The sky-plane position of the companion relative to the primary at each epoch
was computed from the standard Thiele--Innes formulation: Kepler's equation was solved by Newton--Raphson iteration following \citet{danby_fundamentals_1988}.
The angular separation, $\rho$, and position angle, $\theta$, followed from
the sky-projected offsets scaled by the parallax.
The orbital period, eccentricity, argument of periastron, and inferior-conjunction
ephemeris ($T_\mathrm{ic} = $ JD $2\,400\,000 + $ \infConj, when
$\omega_\mathrm{RV}+\nu=90^\circ$) were derived from the RV measurements described
above, while the semi-major axis $a$, inclination $i$, and longitude of the
ascending node $\Omega$ (measured north through east) were adjusted to minimise a
$\chi^2$ sum over the $(\rho_j, \theta_j)$ residuals.
Minimisation used the Nelder--Mead simplex algorithm
as implemented in \texttt{scipy.optimize.minimize} \citep{virtanen_scipy_2020}, launched from a uniform
grid of starting points in $(i_0, \Omega_0)$ to avoid local minima.
From the best-fit $a$ and $i$ we derived the total mass via Kepler's third law, $M_1 + M_2 = 4\pi^2 a^3 / (G P^2)$, and the individual masses by combining
this with the RV mass function
$f(m) = K_1^3 P (1-e^2)^{3/2} / (2\pi G) = M_2^3 \sin^3 i\,/\,(M_1+M_2)^2$,
without any assumption on the mass ratio.
Uncertainties were propagated by Monte Carlo sampling: for each of 20\,000 draws, we independently perturbed all input parameters ($P$, $K_1$, $e$, $\omega$,
$T_\mathrm{ic}$, $\varpi$, and the six astrometric measurements) from Gaussian
distributions defined by their quoted uncertainties, re-ran the full astrometric fit, and recomputed the masses. For each draw, $T_\mathrm{ic}$ was converted to a
periastron time, $T_0$, using the per-sample ($P$, $e$, $\omega$) values. The reported uncertainties correspond to the 16th and 84th percentiles of the
resulting posteriors.
We found \tyc\ with $M_\mathrm{Aa} =$ \MAa\ and \innerwd\ with
$M_\mathrm{Ab} =$ \MAb, yielding a total mass $M_\mathrm{in} =$ \Mtot, with a
semi-major axis $a =$ \smain\ and inclination $i =$ \orbinc\ (Table~\ref{tab:mass_m_dwarf}). The uncertainties
on the individual masses are positively correlated (the error on $M_\mathrm{in}$
exceeds the quadratic sum of the individual uncertainties) because both masses
depend on the same astrometric semi-major axis.
The resulting astrometric fit is shown in Fig.~\ref{fig:kep_fit}; between the first and last \AlopekeH\ epochs (18 December 2023 and 27 September 2024), the separation increased by $\sim$0.07$^{\prime\prime}$ and the position angle by $\sim$3.4$^\circ$.

Because $a_\mathrm{au}$$ = a^{\prime\prime}/\varpi$ at fixed angular semi-major axis $a^{\prime\prime}$ and period, the total mass scales as $M_1+M_2\propto\varpi^{-3}$: a fractional parallax error propagates into three times its own fractional mass error. The formal precision of the adopted parallax ($\varpi_B=$~\BParallaxGDRthree\,mas, 0.01\%) is therefore entirely negligible next to the quoted 0.7--1.7\% mass precisions, which are set by the RV and astrometric terms. The relevant issue is instead a possible systematic difference between the true parallaxes of \tyc\ and \outerwd. Adopting instead the M dwarf's own (flagged-unreliable) \textit{Gaia} DR2 parallax, \mbox{$\varpi_{Aa}=$~\AaParallaxDRtwo} (4.15\% higher than $\varpi_B$), would lower every mass in the system by 11.5\% ($M_\mathrm{in}=1.037\to0.918\,M_\odot$); adopting the midpoint of the two values would lower them by 6.0\%. Both shifts are far larger than the internal precision of the fit, so the choice of parallax matters and deserves justification beyond convenience. A genuine line-of-sight depth offset, of the same order as the $\sim$120\,au projected separation between \tyc\ and \outerwd, is not a viable explanation: at $d=1/\varpi_B\approx\SI{8.46}{\parsec}=1.75\times10^6$\,au, a 120\,au depth offset changes the parallax by only $\sim7\times10^{-5}$ (0.007\%), some three orders of magnitude too small to account for the 4.15\% DR2--DR3 difference, and negligible for the mass (a $2\times10^{-4}$ effect). A much larger true separation is not dynamically excluded: wide binaries of $\sim$1--2\,$M_\odot$ with separations of the order of $10^4$\,au survive perturbations from passing stars and giant molecular clouds for several Gyr up to the age of the Galaxy, while the Galactic tide only truncates bound orbits at $\sim$1\,pc \citep{weinberg_dynamical_1987}, so the true (as opposed to projected) separation between \tyc\ and \outerwd\ could, in principle, be this large, provided the pair is seen nearly along the line of sight. Such a configuration would however require a highly improbable orbital orientation given the small (120\,au) projected separation actually observed, and even in this extreme case the implied depth offset ($\approx10^4$\,au, since the projected separation is negligible by comparison) would shift the parallax by only $\sim0.6\%$ and the mass by $\sim1.7\%$, still far short of the 4.15\%/11.5\% shifts implied by the DR2 parallax discrepancy. The observed discrepancy most likely points to systematics due to the $\sim$10-yr inner-binary orbital wobble imprinted on a single-star, five-parameter \textit{Gaia} astrometric solution, consistent with the extreme \texttt{astrometric\_chi2\_al} already noted in Sect.~\ref{discovery}, rather than from a real difference in distance between the two components of this common-proper-motion, gravitationally bound pair. We therefore adopted $\varpi_B$ as the distance to the system, for which we have no evidence of a systematic error at a level comparable to, let alone exceeding, its quoted statistical precision. A future non-single-star astrometric solution for \tyc\ in \textit{Gaia} DR4 should directly confirm that its parallax is consistent with that of \outerwd.

\subsubsection{Relative geometry of the inner and outer binaries}

\begin{figure}[!ht]
    \centering
    \includegraphics[width=\columnwidth]{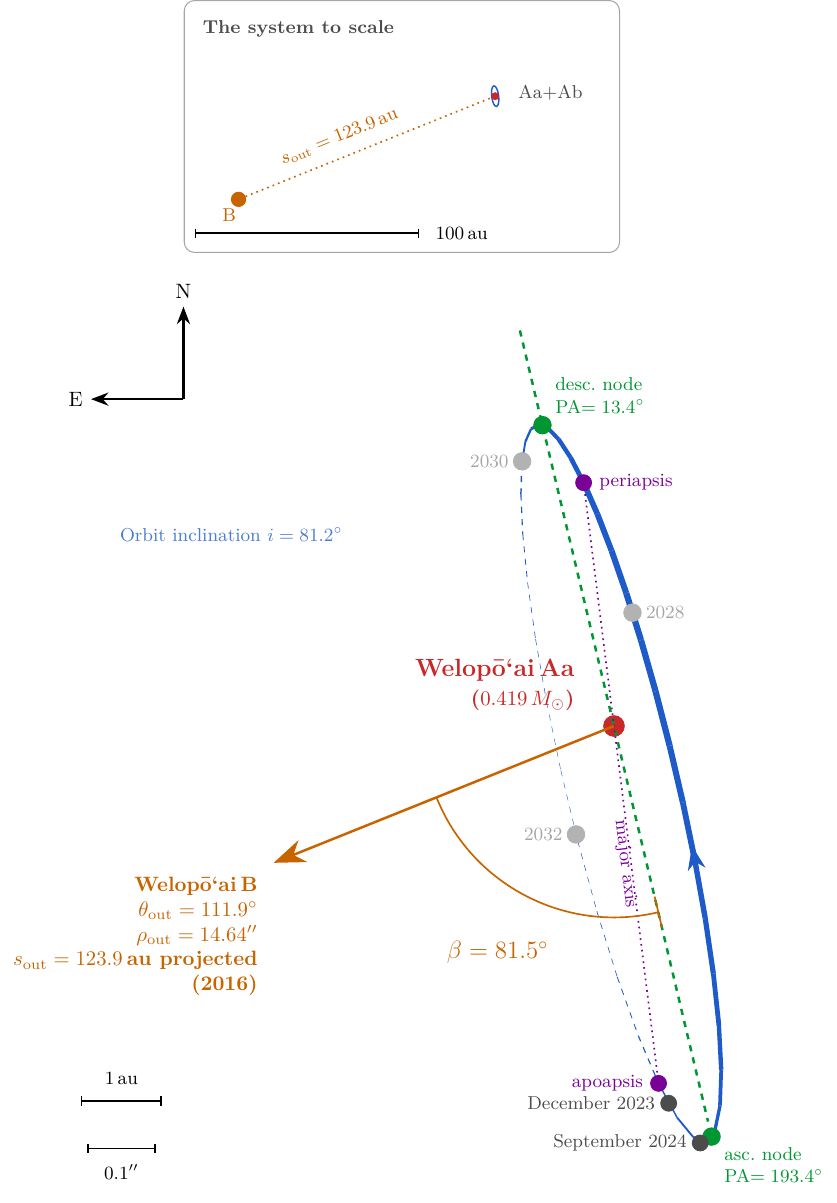}
    \caption{Sky-plane projection of the \innerwd\ orbit around \tyc{} (filled red
      circle at origin). The orbit line width encodes the position of
      \innerwd\ in the direction perpendicular to the plane of sky: thicker blue arcs indicate that \innerwd\ is closer to the observer than \tyc. Dashed arcs mark the part of the orbit farthest from the observer, and the arrowhead gives the direction of motion. Green points and the green dashed line mark the nodes and the line of nodes; purple points mark periapsis and apoapsis, joined by the dotted major axis; dark grey points are the positions measured in December 2023 and September 2024, and light grey points those predicted for 2028, 2030, and 2032. The orange arrow points towards \outerwd, and the orange arc marks $\beta$. The inset at the top shows the whole system to scale: the inner pair (the \tyc--\innerwd\ orbit, too small to resolve at this scale) and \outerwd\ at its projected separation \outerProjSep. The angles and separations of the outer pair are labelled with the symbols of Table~\ref{tab:mass_m_dwarf}: $\theta_{\rm out}$, $\rho_{\rm out}$, $s_{\rm out}$, and the inner--outer misalignment angle $\beta$.}
    \label{fig:sky-projection}
\end{figure}

Having determined the orientation of the inner binary, we measured the sky-plane angle $\beta$ between the line of nodes of the inner orbit and the position angle of the outer companion ($\theta_{\rm out}$$=$~\outerPA; Fig.~\ref{fig:sky-projection}). We found $\beta$~$=$~\ioangle, indicating a significant projected misalignment of the outer companion relative to the inner orbital plane.

\begin{table}[!t]
    \centering
    \caption{Orbital parameters of the inner binary, from SPIRou RVs and \AlopekeH\ astrometry, and the projected properties of the outer binary.}

    \label{tab:mass_m_dwarf}

    \begin{tabular}{ccp{1.55in}}
    \toprule
    \toprule
        Parameter & Best fit  &  Description \\
    \midrule
        \multicolumn{3}{c}{Inner binary} \\
    \midrule
        $M_{\rm Aa}$ ($M_\odot$) & \MAanum & Mass of \tyc \\
        $M_{\rm Ab}$ ($M_\odot$) & \MAbnum & Mass of \innerwd \\
        $P$ (d) & \rvKepPeriodnum & Period \\
        $a_{\rm in}$ (au) & \smainnum  & Semi-major axis\\
        $e$ & \rvKepEcc & Eccentricity \\
        $T_\mathrm{ic}$\tablefootmark{a} (d) & \infConj & Inferior conjunction\\
       $i$ ($^\circ$) & { \orbincnum } &  Orbit inclination to sky plane \\
       $\Omega$ ($^\circ$) & \ascnodeDeg{} & Longitude of ascending node \\
    \midrule
        \multicolumn{3}{c}{Outer binary}\\
    \midrule
    $\rho_{\rm out}$\tablefootmark{b} ($''$) & \outerSeparationnum & Angular separation  \\
    $\theta_{\rm out}$\tablefootmark{b} ($^\circ$) & \outerPAnum & Sky-plane position angle  \\
    $s_{\rm out}$\tablefootmark{c} (au) &  \outerProjSepnum{} & Projected separation  \\
    $P_{\rm out}$ (yr) & \outerPeriodnum{} & \leavevmode Circular orbit, $a_{\rm out}\geq s_{\rm out}$ \\
    $\beta$ ($^\circ$) & \ioanglenum  & \leavevmode Misalignment angle; Fig.~\ref{fig:sky-projection}\\
    \bottomrule
   \end{tabular}
    \tablefoot{Outer binary properties are from \textit{Gaia} astrometry. Detailed Keplerian fit parameters are provided in Table~\ref{tab:ltf_model_posterior}.
    \tablefoottext{a}{JD $-$ 2\,400\,000.}
    \tablefoottext{b}{From \textit{Gaia} DR3 coordinates (2016), consistent with \citet{gonzalezpayo_multiple_2026}.}
    \tablefoottext{c}{$s_{\rm out}=\rho_{\rm out}\,d$, measured from \tyc; the offset of \tyc\ from the barycentre of the inner pair, at most $a_{\rm in}(1+e)M_{\rm Ab}/M_{\rm in}\approx3.3$\,au, dominates over the formal uncertainty.}}
\end{table}

We note the remarkably precise determination of the eccentricity of the inner binary: $e=$~\rvKepEcc. This precision provides a conservative instantaneous constraint on the dynamical state of the system. Implications for Kozai--Lidov (KL) oscillations are discussed in Sect.~\ref{sec:kl}.

\subsection{Properties of \tyc \label{properties_m_dwarf}}
\subsubsection{Photometric temperature, surface gravity, and metallicity}\label{sec:photometricAnalysis}

{
The fundamental stellar parameters ($T_{\mathrm{eff}}$ and radius) were estimated using the empirical calibrations from \citet{mann_how_2015}. These relations connect absolute magnitudes and optical-to-NIR colours to the fundamental properties of K7 to M7 dwarfs. We specifically used their Eq.~(4) with the coefficients of row~1 of their Table~1 for the radius, and their Eq.~(6) with row~10 of their Table~2 ($r-J$ colour and [Fe/H]) for $T_{\mathrm{eff}}$, with the coefficients as corrected by \citet{mann_erratum_2016}, the $r$ magnitude from the American Association of Variable Star Observers (AAVSO) Photometric All-Sky Survey (APASS) DR9 \citep{henden_apass_2016}, the $J$ magnitude from the Two Micron All Sky Survey (2MASS; \citealt{skrutskie_two_2006}), and the photometric [Fe/H] derived below.

The surface gravity was then computed using the following fundamental relation,
\begin{equation}
\log g = \log \left( \left( \frac{M}{M_{\odot}} \right)\left( \frac{R_{\odot}}{R} \right)\right) + \log g_{\odot},
\end{equation}
where $M/M_{\odot}$ and $R/R_{\odot}$ denote the stellar mass and radius in solar units, respectively. For the Sun, we adopted $\log g_{\odot} = 4.4380 \pm 0.0002$ \citep{heiter_gaia_2015}. We obtained for \tyc\,  $T_{\mathrm{eff}} =$ \AaTeffPhot, $R =$ \AaRadius, and $\log g =$ \AaLogG. The stellar mass was derived from Keplerian motion (Sect.~\ref{sec:keplerian_motion}). The reported uncertainties account for both calibration errors and measurement uncertainties, which were propagated within a Monte Carlo framework. Our photometric effective temperature is in agreement with previous $T_\mathrm{eff}$ measurements for \tyc: \AaTeffGaidos\ \citep{gaidos_trumpeting_2014}, \hbox{\AaTeffFouque} \citep{fouque_spirou_2018}, and \hbox{\AaTeffHejazi} \citep{hejazi_chemical_2022}.

The photometric metallicity was then estimated following the empirical method proposed by \citet{duque-arribas_metallicity_2021}, which relates multi-band photometric indices to the iron abundance [Fe/H]. This calibration, based on a Bayesian framework combined with Markov chain Monte Carlo (MCMC) simulations, was trained on 1000 M dwarfs and validated on a further 3\,919 with spectral types up to M5.0\,V (their Eq.~(4) and Table~4), covering the range $-0.45 \leq \mathrm{[Fe/H]} \leq +0.45$ dex with a typical precision of $\sim$0.1 dex. We applied this method to \tyc\ using \textit{Gaia}, 2MASS, and Wide-field Infrared Survey Explorer (\textit{WISE}) photometry together with the parallax of \outerwd, employing the publicly available \texttt{metamorphosis} code\footnote{\url{https://github.com/chrduque/metamorphosis}}, and obtained $\mathrm{[Fe/H]}=$~\AaMHphot\,dex, the mean of the four calibrations of their Table~4 that combine $W1-W2$ and $G_{\rm BP}-G_{\rm RP}$ with $M_G$, $M_J$, $M_H$, or $M_{K_s}$ (\AaMHphotG, \AaMHphotJ, \AaMHphotH, and \AaMHphotK\,dex, respectively).

\subsubsection{Chemical abundances from SPIRou spectroscopy}\label{sec:abundances}

We determined the spectroscopic effective temperature $T_\mathrm{eff}$ and elemental abundances of \tyc\ following the method of \citet{jahandar_comprehensive_2024, jahandar_chemical_2025}, which is optimised for NIR spectroscopy. In brief, the method operates on CFHT/SPIRou spectra and derives the stellar parameters by fitting a set of carefully selected spectral lines, through $\chi^{2}$ minimisation, against PHOENIX-ACES synthetic stellar models \citep{husser_new_2013} convolved to the resolution of SPIRou. We adopted a stellar model grid with a fixed $\log g$ of 4.75, close to the derived \mbox{$\log g=$ \AaLogG} (Sect.~\ref{sec:photometricAnalysis}); small variations ($\pm0.2$~dex) in surface gravity have negligible effect on the results. Our analysis was based on the spectrum built by \texttt{APERO} from all the individual telluric-corrected SPIRou spectra obtained as part of the Keplerian RV monitoring campaign presented in this work. The median S/N of individual spectra is $\sim$100 at 1.6\,$\mu$m, leading to an S/N of $\sim$2000 for the merged template. For each visit, the spectrum was shifted to the stellar rest frame by removing the barycentric velocity and the best-fit Keplerian orbital motion before co-addition.

For the effective temperature, we found a value of \hbox{$T_\mathrm{eff}=$ \AaTeffSpec}, consistent within $1.7\sigma$ with the photometric measurement of $T_\mathrm{eff}=$ \AaTeffPhot\ (Sect.~\ref{sec:photometricAnalysis}). For self-consistency within the analysis of the SPIRou spectra, we used the spectroscopic effective temperature to derive the stellar abundances. For a given $T_{\rm eff}$, individual elemental abundances $[\mathrm{X/H}]$ were measured line-by-line and the scatter among lines was propagated into the per-element uncertainty. To fold the temperature uncertainty in the final abundance error budget, we explored three effective-temperature grid points ($T_{\rm eff} =$ \AaTeffGrid). The resulting abundances are reported in Table~\ref{tab:abundances}, from which we derived a global metallicity $[\mathrm{M/H}] =$ \AaMHspirou, indicating a metal-rich star. The IR spectroscopic [M/H] is in agreement with the value derived from photometry (\AaMHphot~dex, Sect.~\ref{sec:photometricAnalysis}) and measured from high-resolution spectroscopy in the optical (\AaMHFouque~dex; \citealt{fouque_spirou_2018}). 

\subsubsection{Rotation and age}\label{sec:rotation_age_discussion}
The RV observations, analysed with the LBL framework \citep{artigau_line-by-line_2022}, were used to measure temperature changes for \tyc\ over a baseline of \durationRv{} \citep[\textit{dTemp} activity indicator;][]{artigau_measuring_2024}. RV and \textit{dTemp} measurements are provided in Table~\ref{tab:rv_dtemp}. The \textit{dTemp} time series (Fig.~\ref{fig:raw_timeseries}) displays a rotational modulation, which the stellar-activity model of Fig.~\ref{fig:gp_model_timeseries} reproduces. A joint GP analysis of the FWHM and \textit{dTemp} time series, described in Appendix~\ref{sec:methods_intro}, yielded $P_{\rm rot}=$~\AaProtGP, which we adopt as the rotation period of \tyc\ throughout this paper. It agrees with the independent determination from the longitudinal magnetic field ($P_{\rm rot}=$~\BlProt; Sect.~\ref{activity}), and is close to the \residPeriodRot\ signal that remains in the periodograms of the RV, FWHM, and \textit{dTemp} residuals when only the long-term trends and the Keplerian orbit are modelled (Appendix~\ref{sec:methods_ltf}). This period places \tyc\ among the slowly rotating M dwarfs of comparable mass in the MEarth sample, although below the longest periods ($\sim$55--95\,d) measured near 0.4\,$M_\odot$ \citep[][their Fig.~4]{newton_new_2018}. Taken at face value, this is consistent with \tyc\ being a few-Gyr-old field object and in broad agreement with the rotation periods of the coolest (about M3, $T_{\rm eff}\approx3400$--3500\,K) dwarfs of M67 (\SI{4}{\giga\year}) \citep{dungee_4_2022}.

Using the $P_{\rm rot}$--age gyrochronology relation for M2.5--M6.5 dwarfs from
\cite{engle_living_2023} (their Eq.~(4), orthogonal distance regression fit, old track), we derived a
gyrochronological age of \AaGyroAge\ for \tyc. The quoted uncertainty combines
(in quadrature) the formal error from propagating the $P_{\rm rot}$ measurement
uncertainty through Monte Carlo sampling and an intrinsic scatter term of
$\approx\SI{1}{\giga\year}$, comparable to the uncertainty of the age that \citet{engle_living_2023} derive from their relations for the mid-late M dwarfs of M67 ($3.6\pm1.0$\,Gyr; M67 was not used in their calibration). This is broadly consistent with the $\sim$\SI{4}{\giga\year} (3.5--5\,Gyr) age of M67, whose coolest (about M3) dwarfs rotate with similar periods \citep{dungee_4_2022}. However, the cooling age of \innerwd\ alone is $\sim$5.3--5.5\,Gyr, and its total age $\sim$9\,Gyr (Sect.~\ref{properties_white_dwarf}), well above the gyrochronological estimate. If \tyc\ underwent a
common-envelope phase with \innerwd\ during the latter's asymptotic giant branch
evolution, tidal interactions and angular momentum exchange might have spun up the
M~dwarf. The empirical $P_{\rm rot}$--age relations of \cite{engle_living_2023}
are calibrated on cluster members and on M dwarfs in wide binaries (many with WD companions) assumed never to have interacted; they would underestimate the age of a star
whose rotation was altered by binary evolution.

Observational evidence on the rotation of M dwarfs with WD companions is mixed. \citet{rebassamansergas_wddm_2013} found that M dwarfs in wide, never-interacting WD+M-dwarf binaries from SDSS have significantly higher activity fractions than single M dwarfs of the same spectral types, but attributed this mainly to the younger average age of their binary sample rather than to spin-up. Conversely, \GtwoZeroThree, the post-common-envelope system nearest to \welo\ (Sect.~\ref{sec:formation_context}), is not tidally locked despite its $\sim$15-day orbit: its M dwarf has a rotation period that likely exceeds 100\,d (with a tentative 126-d photometric period), which \citet{obrien_wddm_2026} interpreted as the outcome of a brief common-envelope interaction. Rotation is therefore not a simple proxy for the interaction history of such systems. More generally, rotation and activity allow young M dwarfs to be identified but do not yield precise ages \citep{cortescontreras_carmenes_2024}, and given the scatter of the rotation-age relation, a $\sim$47-d period is compatible with a wide range of ages for a mid-M dwarf. The tension between the gyrochronological age of \tyc\ and the ages of the two WDs should therefore not be over-interpreted.

\subsection{Properties of \innerwd \label{properties_white_dwarf}}
\subsubsection{Physical parameters}\label{sec:physParamInnerWD}
Using the photometric data presented in Table \ref{tab:astrometry_photometry}, we performed a photometric fit to obtain the physical parameters of \innerwd\ using the standard method for WDs described at length by \citet{bergeron_chemical_1997}. Since \innerwd\ has no available parallax, we adopted the parallax of \outerwd\ (Sects.~\ref{discovery} and~\ref{sec:keplerian_motion}). Because the available photometric data are scarce and relatively noisy, we additionally constrained the fit to match the dynamical mass ($M_{\rm WD} =$ \MAb, Sect.~\ref{orbital_fit}). The $H$ and $K_s$ band photometric points were excluded from the fit: once $T_{\rm eff}$ is fixed by the optical absolute magnitude and parallax, no pure hydrogen or helium photospheric model can reproduce the observed NIR fluxes at any mass within or beyond the dynamical uncertainty, establishing a NIR colour excess discussed in Sect.~\ref{sec:excess}. The determined parameters for either a hydrogen or a helium-rich WD atmosphere are presented in Table~\ref{tab:system} and the resulting fit is shown in Fig.~\ref{fig:model_wd}; both solutions reproduce the optical photometry well, the pure-H (DA) model marginally better.

\begin{figure}
    \centering
    \includegraphics[width=\linewidth]{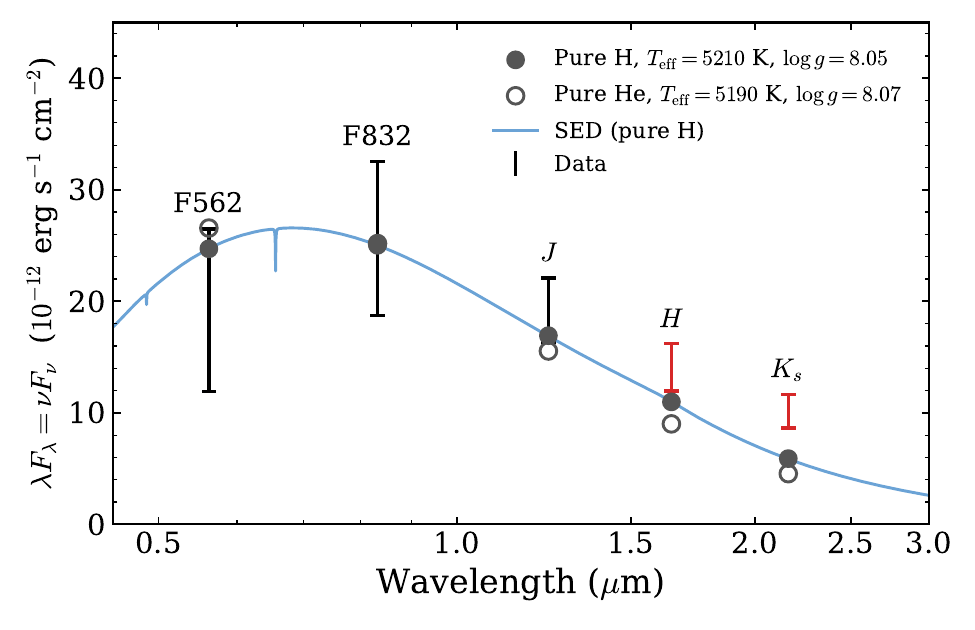}
    \caption{SED of the inner WD: measured photometry (error bars) and band-integrated fluxes of the best-fitting pure-H (filled circles) and pure-He (open circles) models. The blue line is the pure-H model spectrum, whose Balmer lines would allow the spectral type to be confirmed. The $H$ and $K_s$ points (red) are excluded from the fit (Sect.~\ref{sec:excess}).}
    \label{fig:model_wd}
\end{figure}

For both assumed atmospheric compositions (H or He), the effective temperature obtained for \innerwd\ ($T_{\rm eff}$) is near \SI{5200}{\kelvin}, leading to a cooling age of over $\sim$5\,Gyr, broadly consistent with the similar age of \outerwd\ (Table~\ref{tab:system}). Using \texttt{wdwarfdate} \citep{kiman_wdwarfdate_2022}, we inferred that the main sequence progenitor had a mass of $1.44\,M_\odot$, implying a total age of about 9 Gyr for \innerwd. This initial–final mass relation (IFMR) assumes single-star evolution, however. Binary interaction can enhance mass loss and remove the envelope earlier, limiting core growth and producing a lower-mass WD than expected from isolated evolution. In such a scenario, the progenitor mass inferred from the IFMR would be underestimated, leading to an overestimate of the total system age. For example, considering a final mass of about $0.61\,M_\odot$ reached through such enhanced mass loss, from a progenitor more massive than the IFMR implies, the total age of \innerwd\ would instead be approximately 7.8\,Gyr. Such a scenario would be problematic, as the two WDs would no longer have a consistent total age, or one would need to invoke less likely scenarios in which both objects underwent accelerated evolution.

\subsubsection{Spectral type}
The photometric fit slightly favours a hydrogen-rich atmosphere, and we adopted the H-atmosphere solution as our fiducial model for \innerwd\ throughout this paper (Table~\ref{tab:system}); the He-atmosphere solution is retained alongside it as an alternate, since photometry alone does not allow a firm, formal constraint to be placed on the spectral type. Spectroscopic observations are required to confirm the atmospheric composition of \innerwd. At $T_{\rm eff}\sim$\SI{5200}{\kelvin} (nearly independent of assumed composition; Table~\ref{tab:system}), the H$\alpha$ and H$\beta$ Balmer lines should remain detectable, allowing a distinction between hydrogen- and helium-dominated atmospheres (Fig.~\ref{fig:model_wd}). If the WD is helium-rich and not significantly polluted by heavy elements, the spectrum would show no detectable spectral features and the object would therefore be classified as a DC WD. Because the cooling age, progenitor mass, and total age are nearly identical between the two compositions (Table~\ref{tab:system}), this choice does not affect any conclusion drawn elsewhere in the paper.

\subsubsection{IR excess}
\label{sec:excess}
The $H$ and $K_s$ band fluxes of \innerwd\ exceed the predictions of any pure hydrogen or helium photospheric model once the optical photometry and parallax fix $T_{\rm eff}$ (Sect.~\ref{sec:physParamInnerWD}). The $K_s$-band residual corresponds to approximately $3\sigma$ and the $H$-band to $\sim1.5\sigma$, and the discrepancy is robust to the dynamical mass uncertainty. We emphasise that this is a tentative detection: the $H$ and $K_s$ photometry of \innerwd\ could only be anchored to field stars in the GNIRS field of view with published UKIDSS magnitudes, because \tyc\ itself is saturated in these bands (Sect.~\ref{sec:gnirs}). Any zero-point error in the UKIDSS catalogue magnitudes propagates directly into the photometry of \innerwd\ and is not accounted for in the quoted statistical uncertainties; this zero-point term is \gnirsZPerr\ per band (Sect.~\ref{sec:gnirs}); such a systematic could in principle inflate an excess that is smaller, or absent, in reality. Confirming or refuting the excess with photometry referenced to a primary (rather than secondary, UKIDSS-tied) flux standard, or with spectroscopy, should be a priority for follow-up.

This excess could be explained if \innerwd\ hosts a debris disc. For an excess of $\sim 3$\,mJy at $2.15\,\mu$m and a plausible disc temperature of \IRdiskTemp\ \citep[e.g.][]{jura_tidally_2003,jura_infrared_2007}, the disc would have a radius of approximately $9\,R_\oplus$ and a radial width of about $5\,R_\oplus$. Alternatively, a close-in mid-T brown dwarf companion to \innerwd\ would also be compatible with the observations. The nature of this excess remains speculative and warrants further follow-up.

\tyc\ and \innerwd\ together (the unresolved system, historically catalogued as IRAS~21500+5903) have archival MIR photometry, though none of it resolves \tyc\ from \innerwd\ at their $\sim$0\farcs6 separation. Comparing \textit{Gaia}, 2MASS, AllWISE, IRAS, and AKARI photometry of the unresolved system to a spectral-type-matched sample of \IRexcessNmatched\ field M3.0--M4.9 dwarfs from the 20-pc census of \citet{kirkpatrick_initial_2024}, the system matches the field distribution closely: no blended photometric point shows an excess, other than a modest, two-band excess of $\sim$\IRexcessWoneOffset\ and $\sim$\IRexcessWtwoOffset\,mag at WISE $W1$ and $W2$, consistent with the resolved excess reported above towards \innerwd. The one caveat is the archival IRAS 12\,$\mu$m point, which is plausibly a faulty point. See Appendix~\ref{sec:ir_excess_appendix} for further discussion.

\begin{table}[!t]
    \centering
        \footnotesize
    \caption{\label{properties}Properties of the WD components of the system.
    }
    \begin{tabular}{lccc}
    \toprule
    \toprule
    \multicolumn{4}{c}{Physical parameters}\\
    \midrule
    &\multicolumn{2}{c}{\innerwd} & \outerwd \\
    \midrule
    Composition &H & He& H\\
    Spectral type & \ldots & \ldots & DAH\tablefootmark{b}\\
    $T_{\rm eff}$ (K) & \AbTeffHnum & \AbTeffHe & \BTeff\tablefootmark{b}\\
    $\log{g}$ (cgs) &  \AbLogGH & \AbLogGHe & \BLogG\tablefootmark{b}\\
    $M$ ($M_\odot$) & \multicolumn{2}{c}{\MAbnum\,\tablefootmark{a}} & \BMass\tablefootmark{b} \\
    $M_{\rm MS}$\tablefootmark{c} ($M_\odot$) & \AbProgMassH & \AbProgMassHe & \BProgMass\\
    $t_{\rm cool}$\tablefootmark{c} (Gyr) & \AbCoolAgeH & \AbCoolAgeHe & \BCoolAge\\
    $t_{\rm total}$\tablefootmark{c} (Gyr) & \AbTotalAgeH & \AbTotalAgeHe &  \BTotalAge\\
    \bottomrule
    \end{tabular}
    \label{tab:system}
    
    \tablefoot{
    \tablefoottext{a}{From Keplerian fit (dynamical mass, fixed input to the atmospheric fit; Table~\ref{tab:mass_m_dwarf}).}
    \tablefoottext{b}{\cite{caron_spectrophotometric_2023}.}
    \tablefoottext{c}{Calculated using \texttt{wdwarfdate} \citep{kiman_wdwarfdate_2022}; $M_{\rm MS}$ is the main sequence mass of the progenitor.}}
\end{table}

\section{Discussion\label{conclusion}}

The \welo\ system presents a notable hierarchical triple configuration that provides insights into stellar evolution and binary dynamics. In this section, we discuss the tension between the metallicity and the age of the system (Sect.~\ref{section:-metal-age}), its dynamical stability and secular evolution (Sects.~\ref{sec:stability} and~\ref{sec:kl}), the pre-mass-loss configuration of the inner binary (Sect.~\ref{sec:premassloss}), its context among nearby systems (Sect.~\ref{sec:formation_context}), and the follow-up observations that would settle the open questions (Sect.~\ref{sec:followup}). 

\subsection{Metallicity and age}
\label{section:-metal-age}
The supersolar metallicity of \tyc, confirmed by independent methods (i.e. optical and IR high-resolution spectroscopy, photometry), is at odds with the system's age inferred from the total ages of the two WDs, which both point to 9-11\,Gyr, among the oldest stars in the Galaxy. This is not unheard of, as a handful of $>10$\,Gyr stars in Figs.~27 and 28 of \cite{nordstrom_geneva-copenhagen_2004} are seen to have supersolar metallicities. It points to a system in a very sparse part of the age-metallicity diagram, but not an empty one. We considered a few scenarios that could explain this apparent paradox, none of which are particularly convincing:

\begin{itemize}
    \item The system is not co-eval: While the WDs have a consistent age, the M dwarf could be significantly younger, consistent with its high metallicity. This would require capture of the M dwarf, which is very unlikely in a Hubble timescale. In the Galactic disc, encounters within 100\,au, let alone binary capture, occur every $\sim3\times10^{11}$\,yr.
    \item The system is young: WDs in binary systems can have accelerated evolution due to interactions, and binary evolution is known to bias the ages inferred for WDs under single-star assumptions, although for merger products the ages are underestimated rather than overestimated \citep{temmink_looks_2020}. The inner binary is far from compact ($>$100 radii of the main sequence 1.4\,M$_\odot$ progenitor), and the two WDs have consistent total ages, making any scenario that would have both stars undergoing accelerated evolution rather convoluted.
    \item The system is old but the M dwarf is metal-enriched: The metal mass carried by the ejected envelope of the \innerwd\ progenitor ($M_i\approx1.44\,M_\odot$, $M_f=$~\MAb, envelope mass $M_i-M_f\approx0.82\,M_\odot$) scales as $(M_i-M_f)\,Z_\odot\,10^{\mathrm{[Fe/H]}_i}$, assuming it shared the system's initial metallicity. Taking a fiducial pre-pollution metallicity of $\mathrm{[Fe/H]}_i=-0.2$\,dex (close to the mean [Fe/H] of $-0.16$ to $-0.19$\,dex found for field stars aged 2--12\,Gyr in the Geneva-Copenhagen survey; \citealt{nordstrom_geneva-copenhagen_2004}) and the observed spectroscopic value $\mathrm{[M/H]}_f=$~\AaMHspirou\ for \tyc\ ($M_\mathrm{Aa}=$~\MAa; the two metallicity scales are taken as equivalent here), the fraction of the envelope's metals that would need to be captured by the M dwarf, independent of the assumed solar metal fraction $Z_\odot$, is $\approx$69\%: roughly two-thirds of all the metals ever ejected by the WD progenitor, and essentially none of the accompanying hydrogen or helium (which make up $>$98\% of the envelope's mass), would have to land on the M dwarf, otherwise the accreted material would dilute the metallicity back down rather than enhance it. Even under a deliberately generous alternative progenitor mass of 2\,$M_\odot$, a 41\% capture fraction is still required, again essentially hydrogen-free. No established accretion channel (wind Bondi--Hoyle capture, Roche-lobe overflow, or dust-gas segregation) transfers such a large fraction of a giant's ejected metal budget while excluding its hydrogen and helium.
\end{itemize}

Overall, the system may simply be a rare outlier of the age-metallicity distribution. Using simulations from the Besançon Galaxy Model \citep{Robin2003,Czekaj2014,Lagarde2017} of the solar neighbourhood within 100\,pc, we estimated that 5\% of the stars aged from 7 to 12\,Gyr have a supersolar metallicity, [Fe/H] from $+0.05$ to $+0.3$\,dex (and 2\% from $+0.1$ to $+0.2$\,dex).

\subsection{Dynamical stability criteria for hierarchical triple systems}\label{sec:stability}

The long-term stability of the \welo\ system can be assessed using semi-analytical criteria. Following the formalism developed by \citet{mardling_stability_2001}, the stability boundaries for hierarchical systems can be expressed through scaling relations that capture the onset of chaotic behaviour due to resonance overlap. For hierarchical triple systems such as \welo , comprising an inner binary and a distant tertiary companion (i.e., stars are effectively point masses), the stability criterion is
\begin{equation}
\frac{a_{\rm out}}{a_{\rm in}} \simeq C_{\rm tr} \left(\frac{\Omega_{\rm in}}{\psi_p^{\rm out}}\right)^{\alpha}
\label{eq:triple_stability},
\end{equation}
where $a_{\rm in}$ and $a_{\rm out}$ denote the semi-major axes of the inner and outer orbits, respectively, $\Omega_{\rm in}$ is the mean orbital frequency of the inner binary, and $\psi_p^{\rm out}$ represents the orbital frequency of the outer companion at periastron. $C_{\rm tr}$ is an empirically determined constant with a value of 0.013, and $\alpha=4.0$ \citep{mardling_stability_2001}.

This relation can be reformulated as
\begin{equation}
\psi_p^{\rm out} = \rho(e_{\rm out})\Omega_{\rm out},
\end{equation}
where $\Omega_{\rm out}$ is the mean orbital frequency of the outer binary and $\rho(e_{\rm out})$ is the ratio of periastron-to-mean frequencies. From Kepler's laws, we finds that
\begin{equation}
    \rho(e) = \frac{(1+e)^{1/2}}{(1-e)^{3/2}},
\end{equation}
where $e$ is the eccentricity. We note that this converges to 1 for $e=0$ (mean angular velocity equals the periastron value) and diverges to infinity for $e\rightarrow1$. Substituting into Eq.~\eqref{eq:triple_stability}, we obtain
\begin{equation}
\frac{a_{\rm out}}{a_{\rm in}} \simeq C_{\rm tr} \rho(e_{\rm out})^{-\alpha}\left(\frac{\Omega_{\rm in}}{\Omega_{\rm out}}\right)^{\alpha}
\label{eq:stability_rho}.
\end{equation}
From Kepler's third law, we have
\begin{equation}
\frac{\Omega_{\rm in}}{\Omega_{\rm out}} = \left(\frac{a_{\rm out}}{a_{\rm in}}\right)^{3/2} \left(\frac{M_{\rm in}}{M_{\rm tot}}\right)^{1/2},
\end{equation}
where $M_{\rm in}$ is the total mass of the inner binary and $M_{\rm tot}$ is the total mass of the triple system. From the masses given in Tables~\ref{tab:mass_m_dwarf} and \ref{tab:system}, we find
\begin{equation}
\left(\frac{M_{\rm in}}{M_{\rm tot}}\right)^{1/2} \approx0.8.
\end{equation}}
We determined the stability limit to be $\rho(e)\sim17$; the system is stable for $e_{\rm out}\lesssim0.80$, adopting \outerProjSep\ as a proxy for $a_{\rm out}$, although the true boundary could differ if the outer orbit is significantly inclined to the plane of the sky. We note that while the eccentricity of the inner binary is exquisitely constrained, the only constraints concerning the orbit of the outer component of the system are  stability considerations. Configurations with any eccentricity below the above value are stable, regardless of the projection of the orbital inclination of the outer companion.

These semi-analytical results are corroborated by direct dynamical integrations of the triple system over \SI{10}{\mega\year}, which show that it is stable over a broad region of the outer-orbit parameter space. Stability is primarily controlled by the outer pericentre distance, with most configurations having a pericentre distance $\gtrsim$~\outerProjSep\ remaining stable; stable solutions are also found over a wide range of mutual inclinations, indicating that the observed architecture is dynamically compatible with many possible outer-orbit configurations.

\subsection{Kozai--Lidov dynamics}\label{sec:kl}

The presence of the outer WD companion can induce KL oscillations in the inner binary~\citep{kozai_secular_1962,lidov_evolution_1962}. These secular perturbations can cause periodic exchanges between orbital eccentricity and inclination on timescales much longer than the orbital periods. For a triple system with the architecture of \welo, the KL timescale can be estimated as
\begin{equation}
T_{\rm KL} \simeq \frac{2}{3\pi}\frac{P_{\rm out}^2}{P_{\rm in}} \frac{M_{\rm tot}}{M_{\rm out}} \left(1-e_{\rm out}^2\right)^{3/2},
\end{equation}
where $P_{\rm in}$ and $P_{\rm out}$ are the inner and outer periods, respectively, $M_{\rm tot}$ is the total mass of the triple, and $M_{\rm out}$ that of \outerwd\ \citep[][their Eq.~(1)]{fabrycky_shrinking_2007}. Adopting \outerProjSep\ as a lower bound on $a_{\rm out}$, we estimated $T_{\rm KL} \approx$ \KLtimescale{} ($e_\mathrm{out}=0$ and 0.5, respectively); the actual timescale could be longer if the true $a_{\rm out}$ exceeds the projected value ($T_{\rm KL}\propto a_{\rm out}^3$). In either case, $T_{\rm KL}$ is about five orders of magnitude shorter than the WD cooling ages. After the last mass loss episode in the system, most likely from \outerwd\ (its lower mass implies a lower-mass progenitor, $\sim$1.2 versus 1.4\,$M_\odot$ for \innerwd; \outerwd\ therefore left the main sequence later, consistent with its shorter cooling age, Table~\ref{tab:system}), the system has effectively been composed of point-masses considering orbital separations and is therefore unlikely to have undergone eccentricity damping \citep[e.g.][]{eggleton_equilibrium_1998}.

The mutual inclination between the inner and outer orbits is a key parameter for understanding the system's dynamical evolution. As we know the sky-plane orientation of the inner binary, we determined the projected angle $\beta$ between the line of nodes of the inner orbit and the position angle of the outer companion (Fig.~\ref{fig:sky-projection}). The sky-plane angle $\beta$~$=$~\ioangle\ is consistent with a large mutual inclination between the inner and outer orbits; the true three-dimensional mutual inclination remains unconstrained pending an astrometric orbit for the outer pair.

The inner-binary eccentricity ($e =$~\rvKepEcc, precision $\sim$0.8\%) provides a precise snapshot of the current dynamical state; a low eccentricity paired with large mutual inclination is characteristic of a KL turning point, but the mutual inclination is not yet measured. Future constraints on $a_{\rm out}$, $e_{\rm out}$, and the true mutual inclination will enable a direct test of this scenario.

{
\subsection{Pre-mass-loss orbital separation of the inner binary}\label{sec:premassloss}

The inner binary of the system consists of a \MAb\ WD and a \MAa\ M dwarf with  a current orbital semi-major axis of $a_f =$\smain.
The WD is the remnant of an intermediate-mass main sequence star whose initial (zero-age main sequence) mass was $M_{i} \sim1.4$\,M$_\odot$, inferred from the IFMR (Table~\ref{properties}).
As the progenitor evolved off the main sequence and climbed the
asymptotic giant branch (AGB), it shed its envelope in a stellar wind, losing $\sim 0.8$\,M$_\odot$. The current configuration leaves open questions: why KL oscillations during the main sequence phase of the system did not lead to in-spiral, and what prevented orbital decay from the tides raised on the WD progenitor during the AGB phase, both remain unexplained.

To gain insight into the system's main sequence configuration, we computed its orbital separation under the simplistic assumption that the orbital evolution of the inner binary was driven only by mass loss. The total masses of the inner system before and after mass loss are \preMassLossTotal\ and \postMassLossTotal. For an isotropic stellar wind in the adiabatic limit, the semi-major axis evolves \citep{soberman_stability_1997,veras_great_2011} as
\begin{equation}
  \frac{\dot{a}}{a} = -\frac{\dot{M}}{M},
  \label{eq:adiabatic_ode}
\end{equation}
which integrates to the conservation of the invariant
\begin{equation}
  {a\,M = \text{const},}
  \label{eq:aM_const}
\end{equation}
while the orbit expands in inverse proportion to the total mass lost. The pre-mass-loss orbital separation is therefore
\begin{equation}
  a_i = a_f \frac{M_f}{M_i}
      = 4.64\,\textrm{au} \times \frac{1.04}{1.86}
      \approx 2.6\,\textrm{au}.
  \label{eq:ai}
\end{equation}

This value should be compared with the radius at the tip of the AGB: a 1.5\,$M_\odot$ star reaches a maximum bolometric magnitude $M_{\rm bol}=-4.52$ \citep[][their Table~2]{vassiliadis_evolution_1993}, i.e., $L_{\rm bol}\approx5\times10^{3}\,L_\odot$, corresponding to a radius of 1.2\,au for $T_{\rm eff}\approx3000$\,K. Simplistic orbital evolution scenarios point to strong tidal interactions and plausible partial common envelope evolution, complicated by orbital evolution and KL cycles induced by the outer companion. The non-trivial numerical simulations needed to describe such a complex coupled evolution are deferred to future work. 

}

\subsection{Formation history and context among nearby systems}\label{sec:formation_context}

The system's age is well constrained by the cooling ages of both WDs and their progenitor's main sequence lifetime (Table~\ref{tab:system}); the system has a total age of 9--11\,Gyr. At a distance of \sysDistance, \welo\ is the nearest known hierarchical triple system containing two WDs. They share the same total age (i.e., the sum of main sequence lifetime and cooling time), providing a consistency check of their evolutionary histories.

\begin{figure}[!ht]
    \centering
    \includegraphics[width=\linewidth]{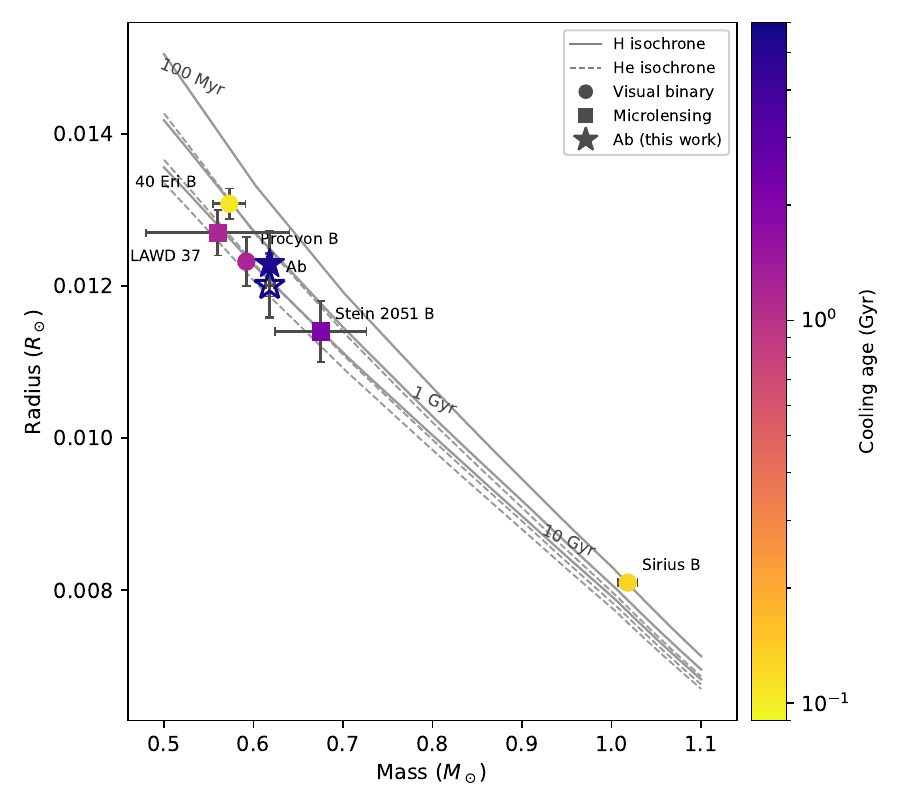}
    \caption{Mass-radius diagram comparing \innerwd\ (star; filled for the adopted H-atmosphere solution, open for the He-atmosphere alternative) to five other nearby WDs with dynamical masses and radii (Table~\ref{tab:mass_radius}): three from visual-binary orbits (circles; Sirius~B, Procyon~B, 40~Eri~B) and two from astrometric microlensing (squares; Stein~2051~B, LAWD~37). All points are coloured by cooling age. Grey lines show pure-hydrogen (solid) and pure-helium (dashed) evolutionary isochrones, labelled at 100\,Myr, 1\,Gyr, and 10\,Gyr, computed from the evolutionary sequences of \citet{bedard_spectral_2020}. }
    \label{fig:mass_radius}
\end{figure}

The evolutionary history of this system is complex and involves at least two mass-loss episodes from progenitor stars. The inner WD, with a mass of \MAb, likely descended from a $\sim$\SI{1.4}{\solarmass} main sequence star based on the IFMR of \citet{cummings_2018}, as implemented in \texttt{wdwarfdate} \citep{kiman_wdwarfdate_2022}. The inner WD mass (0.7\% precision) is among the most accurately determined dynamical WD masses known. For comparison, other nearby dynamical measurements include those of Sirius~B (\massSiriusB;~\citealp{bond_sirius_2017}), Procyon~B (\massProcyonB;~\citealp{bond_procyon_2015}), 40~Eri~B (\masseridB;~\citealp{mason_binary_2017}), Gl~86~B ($0.595\pm0.010\,M_\odot$;~\citealp{brandt_2019}), Stein~2051~B (\SteinMass;~\citealp{sahu_relativistic_2017}), and LAWD~37 (\LawdMass;~\citealp{mcgill_first_2023}), corresponding to 1--14\% fractional errors, the two largest coming from the astrometric-microlensing masses of Stein~2051~B and LAWD~37 rather than from the binary-orbit solutions used for the other four. Such measurements provide critical benchmarks for the WD mass--radius relation \citep{bond_40eri_2017,bedard_massradius_2017}, which underlies the evolutionary models used to derive WD cooling ages \citep[e.g.][]{bedard_spectral_2020}. Figure~\ref{fig:mass_radius} and Table~\ref{tab:mass_radius} place \innerwd\ on the WD mass--radius plane alongside these five benchmarks, using the internally consistent radii and effective temperatures compiled by \citet{bond_40eri_2017} for the three visual-binary systems, and the values reported by \citet{sahu_relativistic_2017} and \citet{mcgill_first_2023} for the two microlensing masses; see also the larger comparison sample of \citet{bedard_massradius_2017}. \innerwd\ significantly extends the cooling-age range probed by this sample: the oldest of the other five WDs, Stein~2051~B, has a cooling age of only \SteinCoolAge\,Gyr, against \AbCoolAgeHnum\,Gyr for \innerwd. Similarly, the M dwarf mass determination (\MAa, 1.7\% precision) is among the most precise for any field M dwarf; the best empirical mass-luminosity relation predicts the masses of single M dwarfs to 2--3\% \citep{mann_how_2019}, and earlier calibrations to 5--10\%, making \tyc\ a valuable calibrator for mass--luminosity and mass--radius relations for low-mass stars, approaching the precision (0.6--1.2\%) of the best-studied eclipsing binary systems~\citep{parsons_scatter_2018}.

\begin{table}[!t]
    \centering
    \caption{Mass-radius comparison of dynamically measured WDs within 20\,pc with a measured radius.}
    \label{tab:mass_radius}
    \footnotesize
    \setlength{\tabcolsep}{1.5pt}
    \begin{tabular}{lcccc}
    \toprule
    \toprule
    Object & $M$ & $R$ & $T_{\rm eff}$ & $t_{\rm cool}$ \\
    & ($M_\odot$) & ($10^{-2}\,R_\odot$) & (K) & (Gyr) \\
    \midrule
    Sirius B & $1.018\pm0.011$ & $0.810\pm0.005$ & $25\,370\pm50$ & \SiriusCoolAge\tablefootmark{b} \\
    Procyon B & $0.592\pm0.006$ & $1.232\pm0.032$ & $7740\pm50$ & \ProcyonCoolAge\tablefootmark{b} \\
    40 Eri B & $0.573\pm0.018$ & $1.308\pm0.020$ & $17\,200\pm110$ & \EridBCoolAge\tablefootmark{b} \\
    Stein 2051 B\tablefootmark{c} & \SteinMassNum & \SteinRadius & \SteinTeff & \SteinCoolAge\tablefootmark{b} \\
    LAWD 37\tablefootmark{d} & \LawdMassNum & \LawdRadius & \LawdTeff & \LawdCoolAge \\
    Ab (H)\tablefootmark{a} & $0.618\pm0.005$ & $1.229\pm0.043$ & $5210\pm320$ & \AbCoolAgeHnum \\
    Ab (He)\tablefootmark{a} & $0.618\pm0.005$ & $1.201\pm0.042$ & $5190\pm340$ & \AbCoolAgeHenum \\
    \bottomrule
    \end{tabular}
    \tablefoot{Sirius B, Procyon B, and 40~Eri~B values compiled by \citet{bond_40eri_2017}; their dynamical masses come from \citet{bond_sirius_2017}, \citet{bond_procyon_2015}, and \citet{mason_binary_2017}, respectively.
    \tablefoottext{a}{Radius of \innerwd\ derived from $M_{\rm Ab}$ and the atmospheric $\log g$ solutions of Table~\ref{tab:system}; not an independent measurement.}
    \tablefoottext{b}{Computed from the same cooling sequences of \citet{bedard_spectral_2020} used for Fig.~\ref{fig:mass_radius} (H-atmosphere tracks for Sirius~B, Procyon~B, and 40~Eri~B; He-atmosphere for Stein~2051~B), for consistency across this table; not independently propagated uncertainties.}
    \tablefoottext{c}{\citet{sahu_relativistic_2017}, via astrometric microlensing.}
    \tablefoottext{d}{\citet{mcgill_first_2023}, via astrometric microlensing; cooling age from the same source.}}
\end{table}

The \periodYears\ orbit of the inner pair, with a semi-major axis of \smain, stands in stark contrast to the hour-to-day periods that characterise the vast majority of post-common-envelope binaries~\citep{zorotovic_post-common-envelope_2010,nebotgomezmoran_post_2011}. As shown in Eq.~\eqref{eq:ai}, the pre-mass-loss inner separation was only a factor of $\sim$2 larger than the maximum AGB radius of the \innerwd\ progenitor, so strong tidal interactions and at least partial envelope grazing were nearly unavoidable. In addition, KL oscillations driven by \outerwd{} must have periodically excited the inner eccentricity, intermittently driving the periastron of \tyc\ towards, and perhaps into, the expanding giant envelope. That the system arrived at its present wide orbit rather than undergoing classical in-spiral~\citep{paczynski_common_1976,ivanova_common_2013} to a tight post-common-envelope configuration indicates that the evolutionary pathway was complex and remains poorly understood. Whether the interaction was merely a partial grazing of the envelope, whether \outerwd\ played a role in halting eccentricity growth at a critical moment, or whether some other mechanism was at work cannot be determined without detailed numerical simulations of the coupled tidal, mass-transfer, and secular dynamical evolution of this hierarchical triple.

{The most directly comparable system in the solar neighbourhood is \GtwoZeroThree\ (GJ~3991) at 7.6\,pc (\textit{Gaia} DR3 parallax), identified as a single-lined spectroscopic binary by \citet{reid_low-mass_1997} and shown by \citet{delfosse_new_1999} to be a post-common-envelope binary consisting of an M3.5\,V primary and a WD companion in a tight \GtwoZeroThreePeriod\ orbit. Its extremely tight orbital separation implies that the M~dwarf likewise survived common-envelope evolution, and together with \tyc\ it forms one of only two such confirmed systems within \SI{10}{\parsec}. Both systems share supersolar metallicity: \GtwoZeroThree\ has [Fe/H]~\GtwoZeroThreeMH\,dex \citep[][a value less secure because its $T_{\rm eff}$ lies below their calibration range]{fouque_spirou_2018}, consistent with that of \tyc, a convergence that is tantalising and deserves investigation with larger samples of post-common-envelope M~dwarf systems. Their orbital configurations, however, differ dramatically: \GtwoZeroThree's 14.7-day period is more than two orders of magnitude shorter than the $\sim$10-year period of the \tyc\ inner pair, suggesting that \tyc\ either experienced a less severe common-envelope phase or underwent substantial post-common-envelope orbital expansion. The WD masses are comparable (\GtwoZeroThree's companion has an inferred mass of \GtwoZeroThreeWDmass; \citealt{delfosse_new_1999}), but \tyc\ offers three key observational advantages: the WD is spatially resolved and directly imaged, giving access to its photometry and, in the future, to its spectrum; the WD mass is directly and precisely determined through the combination of RVs and astrometry rather than inferred from a mass function; and the hierarchical triple configuration with a second WD at $\sim$120\,au provides independent constraints on the system's dynamical history and age that are simply unavailable for \GtwoZeroThree.}

More broadly, the difficulty of finding \innerwd\ in the first place illustrates a persistent incompleteness in the local WD census. \citet{venner_new_2026} recently reported a directly analogous case: an elusive WD companion ($M=0.71^{+0.06}_{-0.05}\,M_\odot$) to the nearby K-dwarf HD~38230, detected by \textit{Gaia} without an astrometric solution and only confirmed via high-contrast imaging spanning several years. At 21\,pc, it is only the 11th-nearest such Sirius-like system known, underscoring how much of the local WD population, particularly faint companions to much brighter primaries, most likely still awaits discovery.

\subsection{Follow-up prospects}\label{sec:followup}
Follow-up observations of the system with existing instrumentation and archival catalogues are possible. The sky-plane acceleration of the outer companion is \outerSkyAccel, implying that the curvature of the orbit can be measured from individual \textit{Gaia} measurements. The relative motion of the M dwarf and outer WD can be measured combining contemporaneous and archival measurements (e.g. Palomar Observatory Sky Survey (POSS) photographic plates from the 1950s; \citealt{minkowski_national_1963}). A \POSSbaseline\ baseline would result in a \POSSpaChange\ change in position angle. Both measurements would add significant constraints on the relative inclination of the inner and outer orbits and shed light on the dynamical evolution of the system.

The tentative $K_s$-band excess of the inner WD may point towards the presence of a debris disc. The system would be an easy target for JWST photometry in the NIR and MIR, shedding light on the nature of that excess.

\section{Conclusions\label{sec:conclusions}}

We present the serendipitous discovery, via a multi-year SPIRou RV monitoring campaign, of a WD companion (\innerwd, $M=$~\MAb) to the nearby M dwarf \tyc, which is itself engaged in a wide, common-proper-motion pair with a second, previously known WD (\outerwd). The resulting hierarchical triple, \welo, is the nearest known system of its kind and has no known analogue within \SI{30}{\parsec}.

A joint Keplerian fit of the SPIRou RVs and \AlopekeH\ relative astrometry yields the inner binary's orbit ($P=$~\periodYearsFull, $a=$~\smain, $e=$~\rvKepEcc) and, independently of any assumption on the mass ratio, both component masses at 0.7--1.7\% precision: $M_\mathrm{Aa}=$~\MAa\ and $M_\mathrm{Ab}=$~\MAb. This makes \innerwd\ one of only a handful of WDs within \SI{20}{\parsec} with a dynamical mass measurement, alongside Sirius~B, Procyon~B, 40~Eri~B, Gl~86~B \citep{brandt_2019}, Stein~2051~B \citep{sahu_relativistic_2017}, and LAWD~37 \citep[][see Sect.~\ref{sec:formation_context}]{mcgill_first_2023}, and its 0.7\% precision is the highest among them; it also makes \tyc\ one of the most precisely measured field M dwarfs.

Both WDs point to a total system age of $\sim$9--11\,Gyr, yet \tyc\ is measurably metal-rich ([M/H]~=~\AaMHspirou), and its rotation period, $P_{\rm rot}=$~\AaProtGP\ (the value we recommend for future use; Sect.~\ref{sec:rotation_age_discussion} and Appendix~\ref{sec:methods_gp}), taken at face value points to a younger gyrochronological age. Reproducing the metallicity would require the M dwarf to have captured roughly two-thirds of the metals, and almost none of the hydrogen or helium, ejected by the inner WD's progenitor, a level of segregated accretion for which no established mechanism is known (Sect.~\ref{section:-metal-age}). The rotation-age tension is milder, as rotation is a poor age indicator for mid-M dwarfs; a binary-interaction spin-up remains possible but is not required (Sect.~\ref{sec:rotation_age_discussion}). These puzzles, together with the still-open question of how the inner binary avoided a classical common-envelope in-spiral (Sect.~\ref{sec:premassloss}), identify \welo\ as a valuable benchmark for future work on binary-star pollution and tidal evolution.

A tentative NIR excess towards \innerwd\ ($3\sigma$ in $K_s$, $1.5\sigma$ in $H$; Sect.~\ref{sec:excess}) is echoed by a modest, band-limited excess at WISE $W1$/$W2$ in the system's unresolved broadband photometry; a discrepant archival IRAS 12\,$\mu$m point, initially suspected of signalling a further MIR excess, instead appears to be a faulty or variable measurement unrelated to circumstellar material (Appendix~\ref{sec:ir_excess_appendix}). Neither the resolved nor the unresolved excess is yet a firm detection: it confirmation would either require a more precise cross-calibration of the underlying photometry, particularly for \innerwd's GNIRS $H$ and $K_s$ measurements, which are currently anchored to secondary (UKIDSS), rather than primary flux standards (Sect.~\ref{sec:excess}). The follow-up avenues summarised in Sect.~\ref{sec:followup} (continued \textit{Gaia} and ground-based astrometry of the outer pair, archival-plate astrometry, and JWST imaging or spectroscopy of \innerwd) would resolve both the 3D architecture of the triple and the nature of the IR excess.

\section*{Data availability}

Tables~\ref{tab:bl} and~\ref{tab:rv_dtemp} are only available in electronic form at the CDS via anonymous ftp to \texttt{cdsarc.u-strasbg.fr} (130.79.128.5) or via \url{http://cdsweb.u-strasbg.fr/cgi-bin/qcat?J/A+A/}.

\begin{acknowledgements}
We thank the referee, Javier Gonz\'alez-Payo, for a careful and constructive report that improved this paper.
{We are grateful to Leslie Ka\okina iu Kimura, Larry Kimura, and Devin Chu of the \okina Imiloa Astronomy Center of Hawai\okina i (University of Hawai\okina i at Hilo), and J.\ Kara Dumaguin of the Canada-France-Hawai\okina i Telescope, for proposing the Hawaiian name \welo\ for this system through the A Hua He Inoa (Calling Forth a Name) effort. The \okina Imiloa Astronomy Center bridges the traditions of Hawaiian navigation and sky knowledge with modern astronomical research conducted on Maunakea, and the name they proposed beautifully captures the orbital dance of this remarkable triple system.}
\'EA, AL, RD, NJC \& CC acknowledge the financial support of the FRQ-NT through the Centre de recherche en astrophysique du Qu\'ebec as well as the support from the Trottier Family Foundation and the Trottier Institute for Research on Exoplanets.
CAP acknowledges support from the Trottier Research Excellence Grants programme, which funded the internship during which this research was conducted. This support provided valuable opportunities for research training through the Trottier Institute for Research on Exoplanets.
\'EA \& RD acknowledge support from the Canada Foundation for Innovation (CFI) programme, the Universit\'e de Montr\'eal and Universit\'e Laval, the Canada Economic Development (CED) programme and the Ministère de l'Économie, Innovation et Énergie (MEIE).
AL acknowledges support from the Fonds de recherche du Qu\'ebec (FRQ) - Secteur Nature et technologies under file \#349961.
This work has been carried out within the framework of the NCCR PlanetS supported by the Swiss National Science Foundation under grants 51NF40\_182901 and 51NF40\_205606.
This research was supported by the Laboratory of Physics and Environmental Chemistry (LPCE) at Universit\'e Joseph Ki-Zerbo in Burkina Faso. It also received financial support through a research mobility grant awarded for a research stay at the Universit\'e de Montr\'eal, within the Trottier Institute for Research on Exoplanets (iREx), thanks to the Canadian government via Global Affairs Canada.
SO is grateful to the Erasmus programme for financial support,
provided through the Universit\'e C\^ote d'Azur, the Universit\'e Joseph Ki-Zerbo,
and the Observatoire de la C\^ote d'Azur / Laboratoire Lagrange in Nice, France.
AKS acknowledges financial support from the Spanish Ministerio de Ciencia, Innovaci\'on y Universidades (MICIU) projects PID2020-117493GB-I00 and PID2023-149982NB-I00, and from La Caixa Foundation (ID 100010434) under the grant LCF/BQ/DI23/11990071.
ACMC acknowledges support from the FCT, Portugal, through the CFisUC project UID/04564/2025 (DOI: 10.54499/UID/04564/2025).
MRS acknowledges support from the research unit FOR2990 (eRO-STEP: Stellar endpoints with eROSITA) funded by the Deutsche Forschungsgemeinschaft.
Some of the observations in this paper made use of the High-Resolution Imaging instrument \AlopekeH. \AlopekeH\ was funded by the NASA Exoplanet Exploration Program and built at the NASA Ames Research Center by Steve B. Howell, Nic Scott, Elliott P. Horch, and Emmett Quigley. \AlopekeH\ was mounted on the Gemini North telescope of the international Gemini Observatory, a programme of NSF’s OIR Lab, which is managed by the Association of Universities for Research in Astronomy (AURA) under a cooperative agreement with the National Science Foundation, on behalf of the Gemini partnership: the National Science Foundation (United States), National Research Council (Canada), Agencia Nacional de Investigaci\'on y Desarrollo (Chile), Ministerio de Ciencia, Tecnolog\'ia e Innovaci\'on (Argentina), Minist\'erio da Ci\^encia, Tecnologia, Inova\c{c}\~oes e Comunica\c{c}\~oes (Brazil), and Korea Astronomy and Space Science Institute (Republic of Korea). This publication makes use of The Data \& Analysis Centre for Exoplanets (DACE), which is a facility based at the University of Geneva (CH) dedicated to extrasolar planets data visualisation, exchange and analysis. DACE is a platform of the Swiss National Centre of Competence in Research (NCCR) PlanetS, federating the Swiss expertise in Exoplanet research. The DACE platform is available at \href{https://dace.unige.ch}{https://dace.unige.ch}.
Based on observations obtained at the Canada-France-Hawai\okina i Telescope (CFHT) which is operated by the National Research Council of Canada, the Institut National des Sciences de l'Univers of the Centre National de la Recherche Scientifique of France, and the University of Hawai\okina i. CFHT is located on Maunakea on Hawai\okina i Island, a mountain of considerable cultural, natural, and ecological significance. Maunakea is a sacred site to Native Hawaiians, also known as K\={a}naka \okina \={O}iwi. Quality observations are made possible by the relentless effort of the entire staff at the Canada-France-Hawai\okina i Telescope. Based on observations obtained with SPIRou, an international project led by Institut de Recherche en Astrophysique et Plan\'etologie, Toulouse, France.
The SPIRou observations were obtained under SLS programme P42 and SPICE programme P45 (PI: J.-F.\ Donati), under run IDs 19BP40, 20AP40, 21AP40, 21BP40, 22AP40 (PI: J.-F.\ Donati), 22BF10, 23AF13, 23AD98, 24AF17 (PI: A.\ Carmona), and 24AC25 (PI: E.\ Artigau). The \AlopekeH\ observations were obtained under Gemini programmes GN-2023-DD-106 and GN-2024B-Q-238 (PI: E.\ Artigau), and the GNIRS imaging under observation GN-2024B-Q-238-59 of that same programme.
This research has made use of the Spanish Virtual Observatory (SVO) Filter Profile Service ``Carlos Rodrigo'', funded by MCIN/AEI/10.13039/501100011033/ through grant PID2023-146210NB-I00.
\end{acknowledgements}

\bibliographystyle{aa}
\bibliography{references}

\begin{appendix}
\counterwithin{table}{section}
\counterwithin{figure}{section}

\section{Origin of the \welo\ name\label{sec:name}}

In Hawaiian, \welo\ combines two roots: \textit{welo}, referring to an inherited or family trait, and \textit{pō\okina ai}, denoting an orbit, circuit, or encircling path. In this cultural framing, \welo\ evokes a `path of family traits', where lineage is traced through motion and where the naming also acknowledges that observations of this system were obtained in Hawai\okina i, on Maunakea.

For this system, the name points not only to an orbital structure, but to an orbital dance as well. Over secular timescales, the outer companion modulates the inner orbit through Kozai--Lidov dynamics, producing a recurring exchange between inclination and eccentricity. Such behaviour is known in hierarchical triples; however, this nearby system offers one of the most compelling examples in the solar vicinity of that long-term gravitational choreography captured by \textit{pō\okina ai}.

\section{Methods: Radial velocity and \textit{dTemp} time series analysis}\label{sec:methods_intro}

We used the modelling framework described by \citet{Stefanov2025a}, hereafter \citetalias{Stefanov2025a}, that is generalised for $N$ datasets (\mbox{$i\in[0..N-1]$}) and $M$ physical quantities (\mbox{$j\in[0..M-1]$}). The \citetalias{Stefanov2025a} framework includes:
(i) long-term functions (LTFs) for data that follow physical or instrumental trends,
(ii) offsets between different datasets,
(iii) dataset-dependent jitters for each physical quantity,
(iv) planetary RV fitting,
(v) stellar activity modelling using Gaussian processes (GPs; \citealp{Rasmussen2006}). GPs are non-parametric models that assume that the correlations between individual measurements can be described in a specific algebraic form \citep{Rasmussen2006}. Stellar activity is often modelled under the assumption that correlations between measurements are sensitive to a certain period -- the stellar-rotation period -- and at the same time, that correlations diminish exponentially at a certain timescale. These two assumptions motivate the squared-exponential periodic (SEP) kernel
(also known as the quasi-periodic, or QP, kernel in the literature), expressed as
\begin{equation}
    k_\text{SEP}(\Delta t) = \kappa^2\exp\left[
    -\frac{\Delta t^2}{2\tau^2}
    -\frac{\sin^2\left(\pi\Delta t/P\right)}{2\eta^2}
    \right],
    \label{eq:sep_kernel}
\end{equation}
where $\kappa^2$ is the maximum amplitude of the kernel, $\tau$ is the timescale of coherence, $P$ is associated with the stellar-rotation period $P_\text{rot}$, and $\eta$ describes the kernel-feature complexity \citep{Haywood2014,rajpaul_gaussian_2015}. We refer to $\eta$ as the `sinescale' (following \citetalias{Stefanov2025a}) on account of its algebraic analogy to the timescale $\tau$. We worked with two approximations of the SEP kernel: the Matérn 3/2 exponential periodic (MEP) kernel and the exponential-sine periodic (ESP) kernel, both introduced by \citet{spleaf2} and implemented in the \texttt{s+leaf}\footnote{\url{https://gitlab.unige.ch/delisle/spleaf}} library \citep{spleaf1,spleaf2}.

We simultaneously fitted GP-based models to the RV and activity-indicator time series. By doing so, we informed the GP kernel of the activity-indicator behaviour of the star, which thereby allows it to better separate planetary from non-planetary signals in RV \citep{rajpaul_gaussian_2015}; a related approach constrains the GP hyperparameters from simultaneous photometry and keeps them fixed in the RV fit \citep{Haywood2014}. In this study, we worked with RV, FWHM, and \textit{dTemp} at the same time. We followed the notation of \citetalias{Stefanov2025a}, and assigned
\mbox{$j=0$} to RV;
\mbox{$j=1$} to FWHM; and
\mbox{$j=2$} to \textit{dTemp}. We followed the $FF'$ formalism, introduced by \citet{Aigrain2012} and extended to a multi-dimensional GP framework by \citet{rajpaul_gaussian_2015}, where one GP kernel is fitted on all physical quantities at the same time; and where for measurements, $y_{i,j}$, at times, $t_{i,j}$, we solved the system of equations, 
\begin{equation}
    \left|\begin{array}{l}
    y_{i,j}=A_j\,G(t_{i,j})+B_j\,(\mathrm{d}G/\mathrm{d}t)_{t=t_{i,j}}\\[0.2ex]
    \shortvdots
    \end{array}\right.\qquad
    \begin{array}{l}
        \forall\, i, \\
        \forall j,
    \end{array}
    \label{eq:multidimensional_system}
\end{equation}
where $A_j$ and $B_j$ are quantity-specific fit hyperparameters, and $G(t_{i,j})$ is the kernel estimation of the stellar activity at given times. We used the \texttt{s+leaf} \texttt{MultiSeriesKernel} procedure to realise this regime.

We utilised nested sampling to evaluate the Bayesian evidence $Z$ of different models numerically, and thereby compare their support by the evidence \citep{Skilling2004}. We used \texttt{ReactiveNestedSampler}, a nested-sampling integrator provided by \texttt{UltraNest} \citep{ultranest}. In every inference instance of $N_\text{param}$ model parameters, we required $40N_\text{param}$ live points and a region-respecting slice sampler that accepted $2N_\text{param}$ steps until the sample was considered independent.

\subsection{Presence of a long-term cycle}\label{sec:methods_ltf}
Long-term trends of instrumental or astrophysical origin must be removed before modelling the stellar-rotation signal, the latter usually lying between a few days and $\sim$\SI{200}{\day} (the \SIrange{5}{200}{\day} search range of \citealt{suarez_mascareno_rotation_2015}). We took all SPIRou measurements in RV, FWHM, and \textit{dTemp}, and composed a 3D model that contained a single elliptical-orbit signal and the following LTF, 
\begin{equation}
    f_j(t_{i,j}) =
    \sum_{n=0}^{p}
    \alpha_{j,n} t^n_{i,j} +
    k_{\text{cyc, }j} \sin\left[\
    \frac{2\pi(t_{i,j}+P_\text{cyc}\,\varphi_{\text{cyc, }j})}{P_\text{cyc}}
    \right],
\end{equation}
where the first term corresponds to a polynomial trend of degree $p$ with coefficients $\alpha_{j,n}$, indexed by quantity and then by order, as in Tables~\ref{tab:ltf_model_posterior} and~\ref{tab:gp_model_posterior}, while the second term introduces a common cycle of amplitude $k_{\text{cyc, }j}$ and phase $\varphi_{\text{cyc, }j}$, with a common cycle period $P_\text{cyc}$. That is to say, we restricted the long-term behaviour of each physical quantity to be described by a different polynomial, but we imposed an additional sine term with a common period in all of them. No GPs were involved in this stage. For the LTF period, we used the prior \mbox{$\mathcal{U}_{\log}\left(200, 1000\right)\,\si{\day}$}.
We found that our best solution in terms of Bayesian evidence is an LTF with a zeroth-order RV trend, a linear FWHM trend, and a linear \textit{dTemp} trend. Table~\ref{tab:ltf_model_posterior} lists all parameters in the model, together with their adopted priors and inferred posteriors.
\begin{table*}
    \centering
    \caption{
        Parameter posteriors of our best model that included only an LTF and \innerwd.
    }
    \label{tab:ltf_model_posterior}
    
    \begin{tabular}{lcccc}
    \toprule
    \toprule
    Parameter name &
    Symbol &
    Unit &
    Prior &
    Posterior \\\midrule
    \multicolumn{5}{c}{LTF parameters} \\
    \noalign{\smallskip}
    Cycle period &
    $P_\text{cyc}$ &
    \si{\day} &
    $\mathcal{U}_{\log}(200,1000)$ &
    \cyclePeriodLTFnum \\
    \noalign{\smallskip}
    Cycle RV phase &
    $\varphi_\text{cyc, 0}$ &
    \ldots &
    $\mathcal{U}(0,1)$\tablefootmark{a} &
    \cyclePhaseRV \\
    \noalign{\smallskip}
    Cycle RV semi-amplitude &
    $k_\text{cyc, 0}$ &
    \si{\metre\per\second} &
    $\mathcal{U}(0,100)$ &
    \cycleAmpRVnum \\
    \noalign{\smallskip}
    Cycle FWHM phase &
    $\varphi_\text{cyc, 1}$ &
    \ldots &
    $\mathcal{U}(0,1)$\tablefootmark{a} &
    \cyclePhaseFWHM \\
    \noalign{\smallskip}
    Cycle FWHM semi-amplitude &
    $k_\text{cyc, 1}$ &
    \si{\metre\per\second} &
    $\mathcal{U}(0,100)$ &
    \cycleAmpFWHMnum \\
    \noalign{\smallskip}
    Cycle differential-temperature phase &
    $\varphi_\text{cyc, 2}$ &
    \ldots &
    $\mathcal{U}(0,1)$\tablefootmark{a} &
    \cyclePhaseDtemp \\
    \noalign{\smallskip}
    Cycle differential-temperature semi-amplitude &
    $k_\text{cyc, 2}$ &
    \si{\kelvin} &
    $\mathcal{U}(0,10)$ &
    \cycleAmpDtempnum \\
    \noalign{\smallskip}
    Systemic velocity\textsuperscript{b} &
    $\alpha_{0,0}$ &
    \si{\metre\per\second} &
    $\mathcal{U}(-10^4,10^4)$ &
    \ltfSysVel \\
    \noalign{\smallskip}
    FWHM first-order correction &
    $\alpha_{1,1}$ &
    \si{\metre\per\second\per\day} &
    $\mathcal{U}(-0.1,0.1)$ &
    \ltfFWHMlinSlope \\
    \noalign{\smallskip}
    FWHM zeroth-order correction &
    $\alpha_{1,0}$ &
    \si{\metre\per\second} &
    $\mathcal{U}(-100,100)$ &
    \ltfFWHMoffset \\
    \noalign{\smallskip}
    Differential-temperature first-order correction &
    $\alpha_{2,1}$ &
    \si{\kelvin\per\day} &
    $\mathcal{U}(-0.1,0.1)$ &
    \ltfDtempLinSlope \\
    \noalign{\smallskip}
    Differential-temperature zeroth-order correction &
    $\alpha_{2,0}$ &
    \si{\kelvin} &
    $\mathcal{U}(-100,100)$ &
    \ltfDtempOffset \\
    \noalign{\smallskip}
    \multicolumn{5}{c}{Dataset parameters} \\
    \noalign{\smallskip}
    SPIRou RV jitter &
    $J_{0,0}$ &
    \si{\metre\per\second} &
    $\mathcal{U}(0,100)$ &
    \ltfRVjitter \\
    \noalign{\smallskip}
    SPIRou FWHM jitter &
    $J_{0,1}$ &
    \si{\metre\per\second} &
    $\mathcal{U}(0,100)$ &
    \ltfFWHMjitter \\
    \noalign{\smallskip}
    SPIRou differential-temperature jitter &
    $J_{0,2}$ &
    \si{\kelvin} &
    $\mathcal{U}(0,10)$ &
    \ltfDtempJitter \\
    \noalign{\smallskip}
    \multicolumn{5}{c}{\innerwd} \\
    \noalign{\smallskip}
    Period &
    $P$ &
    \si{\day} &
    $\mathcal{U}_{\log}(10^3,10^4)$ &
    \rvKepPeriodnum \\
    \noalign{\smallskip}
    Phase &
    $\varphi$ &
    \ldots &
    $\mathcal{U}(0,1)$\tablefootmark{a} &
    \rvKepPhase \\
    \noalign{\smallskip}
    Inferior-conjunction ephemeris &
    $\varepsilon$ &
    JD$-$\num{2\,400\,000} &
    derived &
    \infConj \\
    \noalign{\smallskip}
    RV semi-amplitude &
    $k_\text{rv}$ &
    \si{\metre\per\second} &
    $\mathcal{U}_{\log}(10^3,10^5)$ &
    \rvKepAmpnum \\
    \noalign{\smallskip}
    Eccentricity &
    $e$ &
    \ldots &
    $\mathcal{U}(0,1)$ &
    \rvKepEcc \\
    \noalign{\smallskip}

    Argument of periastron &
    $\omega$ &
    \si{\radian} &
    $\mathcal{U}(0,2\pi)$\tablefootmark{a} &
    \rvKepOmega \\
    \noalign{\smallskip}
\bottomrule
    \end{tabular}
    \tablefoot{Uncertainties are the 16th and 84th percentiles of the posterior distributions.
    \tablefoottext{a}{Wrapped parameter.}
    \tablefoottext{b}{While the formal errors on systemic velocity are at the \SI{10}{\metre\per\second} level, the micro-physics-related uncertainties lead to $\sim$\SI{40}{\metre\per\second} uncertainties on the true line-of-sight RV (37\,m\,s$^{-1}$ template-to-template dispersion; \citealt{artigau_line-by-line_2022}), before line-position and gravitational-redshift systematics are considered.}
    }
\end{table*}
We measured a cycle period of \cyclePeriodLTF\ with a well-behaving, unimodal posterior shape. We obtained the following RV, FWHM and \textit{dTemp} semi-amplitudes of the cycle:
\cycleAmpRV,
\cycleAmpFWHM, and
\cycleAmpDtemp\ respectively. These results underline a significant detection of the activity signal in the  RV sequence. The cycle phase somewhat differs for different quantities -- we obtained RV, FWHM, and \textit{dTemp} phases of
\cyclePhaseRV,
\cyclePhaseFWHM, and
\cyclePhaseDtemp\ respectively. FWHM and \textit{dTemp} appear to be in phase within $1\sigma$, and at the same time, both of them are out of phase relative to RV. The RV elliptic-orbit signal is constrained remarkably well, with a period of
\rvKepPeriod, a semi-amplitude of
\rvKepAmp, and an eccentricity of
\rvKepEcc.

\begin{figure*}[!htbp]
    \centering
    \includegraphics[width=\linewidth]{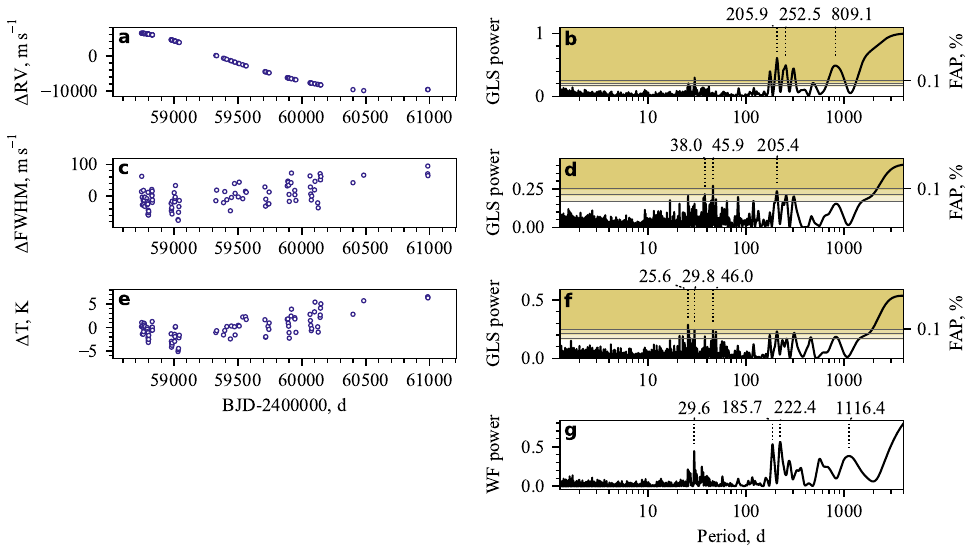}
    \caption{
    Raw \tyc\ time series (blue markers) of mean-subtracted:
    (a) RV;
    (c) FWHM;
    (e) \textit{dTemp};
    (b,d,f) Associated wide-period generalised Lomb--Scargle periodograms (GLSPs). Three FAP levels, 10\%, 1\%, and 0.1\%, split GLSP ordinates in bands of different colour; vertical dashed lines mark the highest peaks, labelled with their periods in days; (g) Window function of the time series.
    }
    \label{fig:raw_timeseries}
\end{figure*}

\begin{figure*}[!htbp]
    \centering
    \includegraphics[width=\linewidth]{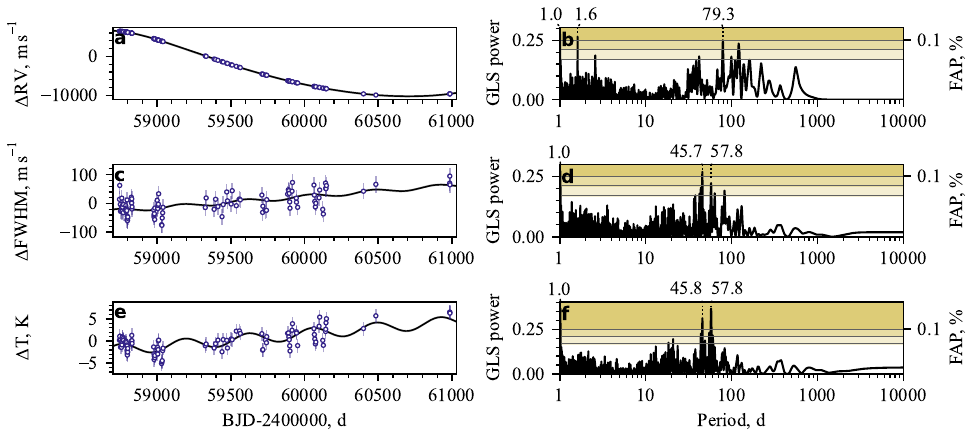}
    \caption{
    Same as Fig.~\ref{fig:raw_timeseries}, but for the time series against our LTF model that accounts for \innerwd\ (black): (a) RV, (c) FWHM and (e) \textit{dTemp}, with (b,d,f) the GLSPs of the residuals, accounting for the model jitter.
    }
    \label{fig:ltf_model_timeseries}
\end{figure*}

Figure~\ref{fig:ltf_model_timeseries} shows this baseline-model fit against data, as well as the periodograms of residuals, having accounted for the model jitter. The modelling of the long-term cycle and the RV signal was enough for no peak beyond \SI{200}{\day} to remain significant at the 10\% false alarm probability (FAP; \citealp{Baluev2008}) level. On the short-period end, we saw that a \residPeriodRot\ signal remained significant in RV, FWHM, and \textit{dTemp}. There is an additional signal at \residPeriodRV\ in RV, as well as a \residPeriodFWHM\ signal in both FWHM and \textit{dTemp}. The latter would be explained by a \residPeriodAlias\ alias in the raw time series that we could identify in the window function (Fig.~\ref{fig:raw_timeseries}g). The \SI{163}{\day} RV peak is not an alias of the signals discussed here.

\subsection{Presence of short-term stellar activity from \tyc}\label{sec:methods_gp}

Gaussian processes are known to absorb periodicities larger than $P_\text{rot}$ with ease. This in itself justifies the modelling of longer-term behaviours by an LTF, before a GP is introduced. Nevertheless, GPs can absorb part or all of a long-term trend: \citetalias{Stefanov2025a} found that the evidence no longer favoured a sinusoidal LTF term once a GP was included, which they attributed to the kernel fitting the long-term trend, and in GJ~341 a quasi-periodic GP alone captured a $\sim$4000\,d magnetic cycle \citep{DiTomasso2025}. We were unable to model short-term stellar activity RV variation through any GP kernel, neither in one- nor in multidimensional mode. We attribute this to the large variation in RV that goes beyond \SI{e4}{\metre\per\second} (Fig.~\ref{fig:raw_timeseries}), far beyond the typical use case of GPs to model stellar activity (commonly  under \SI{e2}{\metre\per\second}). We attempted to correct the RV contribution from \innerwd\ first, then use these modified time series to model for the LTF and stellar activity. This endeavour proved fruitless. We reasoned that there is likely a higher order remnant after the RV modelling of \innerwd\ which prevents us from modelling stellar activity in RV.  Alternatively, \tyc\ itself may exhibit minute stellar activity in RV relative to instrument precision.

We took the two activity indicators from our aforementioned LTF model: FWHM and \textit{dTemp}, and then augmented it with a GP stellar-activity component. Additionally, we constrained the common-cycle period to the more informed 
\mbox{$\mathcal{U}(300,500)\,\si{\day}$}, having gained insights from Sect.~\ref{sec:methods_ltf}. In terms of  kernel selection, we tested for the MEP kernel, as well as the ESP kernel with two, three or four harmonics (hereafter ESP2, ESP3, ESP4). The FWHM and the \textit{dTemp} time series appear remarkably similar to one another (Fig.~\ref{fig:raw_timeseries}c,e), which prompted us to assume no gradients of one relative to the other. We set $B_1,B_2=0$ in Eq.~\eqref{eq:multidimensional_system} and excluded them from sampling. We used uninformed priors for the rotation period, the timescale and the sinescale:
\mbox{$P_\text{rot}=\mathcal{U}_{\log}(2,200)\,\si{\day}$},
\mbox{$\tau=\mathcal{U}_{\log}(20,10^4)\,\si{\day}$}, and
\mbox{$\eta=\mathcal{U}(0,3)$}. We reported the following log-Bayesian evidences $\ln Z$ for the SEP kernels we tested for:
\lnZmep\ for MEP,
\lnZespTwo\ for ESP2,
\lnZespThree\ for ESP3, and
\lnZespFour\ for ESP4. In other words, $\ln Z$ increased monotonically with the complexity of kernels, but not with an overwhelming preference towards a specific model (which is conventionally implied by \mbox{$\Delta\ln Z>5$}). We chose to adopt the MEP kernel in our analysis, by virtue of being the most algebraically simple, as well as faring best in terms of $\ln Z$.

\begin{figure*}[!htbp]
    \centering
    \includegraphics[width=\linewidth]{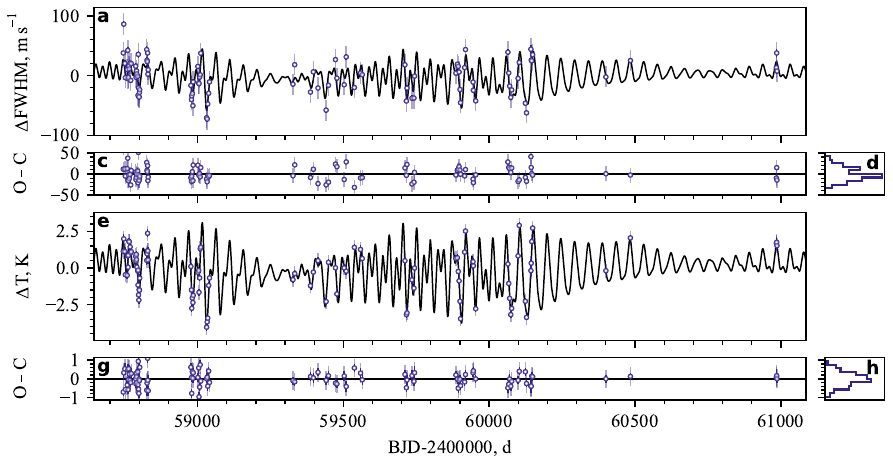}
    \caption{GP stellar activity model fitted to the FWHM and \textit{dTemp} time series of \tyc. (a) FWHM and (e) \textit{dTemp} measurements (blue circles) with the GP model (black line); (c, g) residuals (O$-$C) and (d, h) their distributions.}
    \label{fig:gp_model_timeseries}
\end{figure*}

Figure~\ref{fig:gp_model_timeseries} shows the results of our stellar-activity model in FWHM and \textit{dTemp}. We observed good agreement between model and data, both along the full baseline and at intervals comparable to $P_\text{rot}$. The robustness of our model is supported by the distributions of the FWHM and \textit{dTemp} residuals (Fig.~\ref{fig:gp_model_timeseries}d,h) -- we found no apparent trends in the O-C diagrams with time, and they appear to follow a Gaussian-like distribution. The GLSPs of residuals show no other significant peaks beyond 1\% FAP.

\begin{table*}
    \centering
    \caption{
        Parameter posteriors of our multi-dimensional GP model in FWHM and differential temperature.
    }
    \label{tab:gp_model_posterior}
    
    \begin{tabular}{lcccc}
    \toprule
    \toprule
    Parameter name &
    Symbol &
    Unit &
    Prior &
    Posterior \\\midrule
    \multicolumn{5}{c}{LTF parameters} \\
    \noalign{\smallskip}
    Cycle period &
    $P_\text{cyc}$ &
    \si{\day} &
    $\mathcal{U}(300,500)$ &
    \cyclePeriodGPnum \\
    \noalign{\smallskip}
    Cycle FWHM phase &
    $\varphi_\text{cyc, 1}$ &
    \ldots &
    $\mathcal{U}(0,1)$\tablefootmark{a} &
    \cyclePhaseFWHMgp \\
    \noalign{\smallskip}
    Cycle FWHM semi-amplitude &
    $k_\text{cyc, 1}$ &
    \si{\metre\per\second} &
    $\mathcal{U}(0,100)$ &
    \cycleAmpFWHMgpnum \\
    \noalign{\smallskip}
    Cycle differential-temperature phase &
    $\varphi_\text{cyc, 2}$ &
    \ldots &
    $\mathcal{U}(0,1)$\tablefootmark{a} &
    \cyclePhaseDtempGP \\
    \noalign{\smallskip}
    Cycle differential-temperature semi-amplitude &
    $k_\text{cyc, 2}$ &
    \si{\kelvin} &
    $\mathcal{U}(0,10)$ &
    \cycleAmpDtempGPnum \\
    \noalign{\smallskip}
    FWHM first-order correction &
    $\alpha_{1,1}$ &
    \si{\metre\per\second\per\day} &
    $\mathcal{U}(-0.1,0.1)$ &
    \gpFWHMlinSlope \\
    \noalign{\smallskip}
    FWHM zeroth-order correction &
    $\alpha_{1,0}$ &
    \si{\metre\per\second} &
    $\mathcal{U}(-100,100)$ &
    \gpFWHMoffset \\
    \noalign{\smallskip}
    Differential-temperature first-order correction &
    $\alpha_{2,1}$ &
    \si{\kelvin\per\day} &
    $\mathcal{U}(-0.1,0.1)$ &
    \gpDtempLinSlope \\
    \noalign{\smallskip}
    Differential-temperature zeroth-order correction &
    $\alpha_{2,0}$ &
    \si{\kelvin} &
    $\mathcal{U}(-100,100)$ &
    \gpDtempOffset \\
    \noalign{\smallskip}
    \multicolumn{5}{c}{Dataset parameters} \\
    \noalign{\smallskip}
    SPIRou FWHM jitter &
    $J_{0,1}$ &
    \si{\metre\per\second} &
    $\mathcal{U}(0,100)$ &
    \GPjitterFWHMnum \\
    \noalign{\smallskip}
    SPIRou differential-temperature jitter &
    $J_{0,2}$ &
    \si{\kelvin} &
    $\mathcal{U}(0,10)$ &
    \GPjitterDtempnum \\
    \noalign{\smallskip}
    \multicolumn{5}{c}{Stellar-activity parameters} \\
    \noalign{\smallskip}
    Timescale &
    $\tau$ &
    \si{\day} &
    $\mathcal{U}_{\log}(20, 10^{4})$ &
    \GPtaunum \\
    \noalign{\smallskip}
    Period &
    $P_\text{rot}$ &
    \si{\day} &
    $\mathcal{U}_{\log}(2,200)$ &
    \AaProtGPnum \\
    \noalign{\smallskip}
    Sinescale (harmonic complexity) &
    $\eta$ &
    \ldots &
    $\mathcal{U}(0,3)$ &
    \GPeta \\
    \noalign{\smallskip}
    FWHM amplitude &
    $A_1$ &
    \si{\metre\per\second} &
    $\mathcal{U}_{\log}(0.01,100)$ &
    \GPampFWHMnum \\
    \noalign{\smallskip}
    Differential-temperature amplitude &
    $A_2$ &
    \si{\kelvin} &
    $\mathcal{U}_{\log}(0.01,100)$ &
    \GPampDtempnum \\
    \bottomrule
    \end{tabular}
    \tablefoot{Uncertainties are the 16th and 84th percentiles of the posterior distributions.
    \tablefoottext{a}{Wrapped parameter.}
    }
\end{table*}

Table~\ref{tab:gp_model_posterior} displays the posteriors of our stellar-activity model. Having accounted for stellar activity, we obtained an LTF cycle period of
\mbox{\cyclePeriodGP}, with an FWHM semi-amplitude of
\mbox{\cycleAmpFWHMgp}, and a \textit{dTemp} semi-amplitude of
\mbox{\cycleAmpDtempGP}. Although we retrieved the same common-cycle period compared to the LTF-only model
(\mbox{\cyclePeriodLTF}; Table~\ref{tab:ltf_model_posterior}), we remark that this time the \textit{dTemp} semi-amplitude is significantly lower, while the FWHM amplitude agrees within its uncertainty (against
\mbox{\cycleAmpFWHM} and 
\mbox{\cycleAmpDtemp}; Table~\ref{tab:ltf_model_posterior}). We obtained the following cycle phases:
\mbox{\cyclePhaseFWHMgp} in FWHM and
\mbox{\cyclePhaseDtempGP} in \textit{dTemp}. Our model assigned a jitter of
\mbox{\GPjitterFWHM} in FWHM, and
\mbox{\GPjitterDtemp} in \textit{dTemp}.

We constrained the stellar-rotation period of \tyc\ to
\mbox{\AaProtGP}, and reported \mbox{$\tau=$ \GPtau}. The ratio $\tau/P_\text{rot}$ is larger than unity, which likely implies a light-curve morphology that is not Sun-like \citep{Giles2017}. That being said, we emphasise that such assessments must be subject to scrutiny, since MEP models tend to return  larger timescales than their ESP counterparts (e.g. \citealp{Stefanov2025b,Stoeva2026}). We obtained a sinescale of \GPeta\ and the following GP amplitudes:
\mbox{$A_1=$ \GPampFWHM}, and
\mbox{$A_2=$ \GPampDtemp}. Both amplitudes are positive, and imply that FWHM grows with
\mbox{\fwhmDtempSlope}\ for every additional kelvin in differential temperature. A dark spot on the approaching/receding half of the stellar disc would suppress the blue/red-shifted half of a line profile, which nudges said half closer to the core, thereby decreasing the FWHM reading. At the same time, a dark spot on the stellar surface would lower the disc-average temperature; meaning that \textit{dTemp} would fall together with the decrease of FWHM. Therefore, a positive correlation between FWHM and \textit{dTemp} implies dark spots on the stellar surface of \tyc.

We ran a variant of the detection-limit test in \citet{Stefanov2025b}. We took our best model that included an LTF and \innerwd\ (Fig.~\ref{fig:ltf_model_timeseries}), and augmented it with an extra sinusoid in RV. We chose a period interval of
\mbox{$\mathcal{U}_\text{log}(1,1000)\,\si{\day}$}
and split it in 100 bins. For each of these bins, we defined a separate `bin model' that took its bin edges as the period prior of the sine, and an RV semi-amplitude of $\mathcal{U}(0,30)\,\si{\metre\per\second}$. Then, we sampled only for the period of the sine, its RV semi-amplitude, as well as the RV zeroth-order correction and jitter following priors of the original model (Table~\ref{tab:ltf_model_posterior}). The remaining parameters were kept fixed at the posterior medians of the original model. Finally, for each bin, we considered the $\pm 3\sigma$ percentiles of the RV semi-amplitude posterior of the augmented sine. The $+3\sigma$ percentile can be interpreted as the maximum semi-amplitude a signal would have, if it was significant whatsoever, and we took this value as our detection limit. The $-3\sigma$ percentile tells of any significant signals that stand well away from the zero and that could deserve attention in further studies.

\begin{figure}
    \centering
    \includegraphics[width=\linewidth]{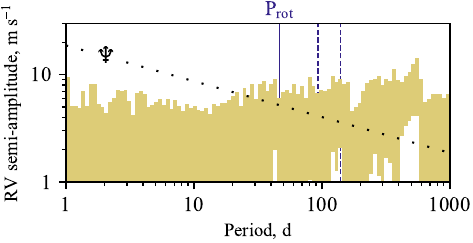}
    \caption{
    Detection limits of \tyc. Shaded bands show the $3\sigma$ RV semi-amplitude posterior of a modelled planet in our best model, in 100 log-spaced orbital-period bins. A black dotted line shows the RV semi-amplitudes that a Neptune-mass planet would inject. Our measured stellar rotational period $P_\text{rot}$ is marked with a solid blue line, and its first multiples with dashed blue lines.
    }
    \label{fig:detection_limits}
\end{figure}

Figure~\ref{fig:detection_limits} displays the result of this exercise. We note the presence of bin models with somewhat separated semi-amplitude posteriors near our measured stellar rotational period, $P_\text{rot}$, its first multiples, and at periods beyond \SI{200}{\day}. We stress that the inability to model stellar activity in RVs prevents us from determining the nature of these signals. Further works would benefit from intensive short-term RV monitoring to first establish an activity model in RV and then seek planetary candidates. Nevertheless, we ruled out all RV signals with periods shorter than \SI{20}{\day} that would correspond to Neptune-mass or more massive planets.

\section{Astrometric and photometric properties of \welo\label{sec:astrometric_table}}
{Table~\ref{tab:astrometry_photometry} summarises the astrometric and photometric properties of all three components of the \welo\ system, drawn from \textit{Gaia}~DR3, 2MASS, APASS, Panoramic Survey Telescope and Rapid Response System (Pan-STARRS), and our own \AlopekeH\ and GNIRS measurements.}
\begin{table*}[!ht]
    \centering
    \caption{Astrometric and photometric properties of the \welo\ system.}
    \begin{tabularx}{\linewidth}{XXXX}
    \toprule
    \toprule
       Property & \innerwd & \tyc & \outerwd \\
       \textit{Gaia} DR3 ID & \ldots & \AaGaiaDRthreeID & \BGaiaDRthreeID\\
    \midrule
    \multicolumn{4}{c}{Astrometric properties} \\
    \midrule
    $\alpha$ (ICRS, ep. 2016.0) & \ldots &\AaRA\tablefootmark{a} & \BRA\tablefootmark{b}\\
    $\delta$ (ICRS, ep. 2016.0) & \ldots &\AaDec\tablefootmark{a} & \BDec\tablefootmark{b}\\
    $\mu_\alpha\cos\delta$ (\si{\mas\per\year}) & \ldots &\AaMuAlpha\tablefootmark{a} & \BMuAlpha\tablefootmark{b} \\
    $\mu_\delta$ (\si{\mas\per\year}) & \ldots &\AaMuDelta\tablefootmark{a} &\BMuDelta\tablefootmark{b} \\
    $\varpi$ (mas) &\ldots&\ldots& \BParallaxGDRthree\tablefootmark{b} \\
    $d$ (pc) &\ldots& \ldots& \sysDistanceFull\tablefootmark{b} \\
    \midrule
    \multicolumn{4}{c}{Photometric properties (mag)}\\
    \midrule
    $g$ & \ldots & \ldots & \Bgmag\tablefootmark{d} \\
    $G_{\rm BP}$ {\tablefootmark{c}} & \ldots & \AaGBPmag & \BGBPmag\tablefootmark{b}\\
    $r$ & \ldots & \ldots & \Brmag\tablefootmark{d} \\
    $F562$ {\tablefootmark{c}} & \AbFiveSixTwomag\tablefootmark{e} & \AaFiveSixTwomag & \ldots \\
    $i$ & \ldots & \ldots & \Bimag\tablefootmark{d} \\
    $G_{\rm RP}$ {\tablefootmark{c}} & \ldots & \AaGRPmag & \BGRPmag\tablefootmark{b}\\
    $F832$ {\tablefootmark{c}} & \AbEightThreeTwomag\tablefootmark{e} & \AaEightThreeTwomag & \ldots \\
    $z$ & \ldots & \ldots & \Bzmag\tablefootmark{d} \\
    $y$ & \ldots & \ldots & \Bymag\tablefootmark{d} \\
    $J$ & \AaJmag\tablefootmark{e} & \TycJmag\tablefootmark{f} & \BJmag\tablefootmark{f}\\
    $H$ & \AaHmag\tablefootmark{e} & \TycHmag\tablefootmark{f} & \BHmag\tablefootmark{f}\\
    $K_{\rm s}$ & \AaKmag\tablefootmark{e} & \TycKmag\tablefootmark{f} & \BKmag\tablefootmark{f}\\
    \bottomrule
    \end{tabularx}
    \label{tab:astrometry_photometry}
    \tablefoot{
    \tablefoottext{a}{\textit{Gaia} DR3 position at epoch 2016.0, which does not take the orbital motion of \tyc\ into account explicitly, and \textit{Gaia} DR2 proper motion: \tyc\ has only a two-parameter solution in DR3, with neither proper motion nor parallax (Sects.~\ref{discovery} and~\ref{sec:keplerian_motion}).}
    \tablefoottext{b}{\textit{Gaia} DR3 \citep{gaia_collaboration_dr3_2023}; $d=1/\varpi$; magnitude uncertainties include a 0.010\,mag calibration floor.}
    \tablefoottext{c}{Non-standard filters (FWHM ranges): $F562$ = \AaFilterBluebandpass\,nm, $F716$ = \AaFilterSevenOneSixbandpass\,nm, $F832$ = \AaFilterRedbandpass\,nm; $G_{\rm BP}$ and $G_{\rm RP}$ are the \textit{Gaia} DR3 passbands (Pan-STARRS $g,r,i,z,y$ and 2MASS $J,H,K_{\rm s}$ bandpasses are standard and not repeated here). Full filter transmission curves for all bandpasses listed in this table are available through the Spanish Virtual Observatory (SVO) Filter Profile Service (\url{http://svo2.cab.inta-csic.es/theory/fps/}; \citealt{rodrigo_svo_2012,rodrigo_svo_2020,rodrigo_svo_2024}).}
    \tablefoottext{d}{\cite{caron_spectrophotometric_2023}.}
    \tablefoottext{e}{From \AlopekeH\ contrast measurements (optical) and GNIRS imaging (NIR).}
    \tablefoottext{f}{\cite{skrutskie_two_2006}.}}
\end{table*}

\section{Complementary tables: longitudinal magnetic field, elemental abundances, and RV time series}\label{timeseries}

This appendix provides the full longitudinal magnetic field, elemental abundances, and RV data (Sects.~\ref{magfield} and \ref{sec:abundances}, and Appendix~\ref{sec:methods_intro}) in tabular form. Table~\ref{tab:bl} lists the \nStokesVkept\ longitudinal magnetic field measurements (Sect.~\ref{magfield}) and Table~\ref{tab:abundances} the elemental abundances (Sect.~\ref{sec:abundances}). Table~\ref{tab:rv_dtemp} lists the \nRVframes\ SPIRou barycentric RVs and
differential temperature indicators (\textit{dTemp}, $\Delta T_{3500}$; \citealt{artigau_measuring_2024}) for \tyc\ obtained with \texttt{APERO} \citep{cook_apero_2022} using the \texttt{LBL} framework \citep{artigau_line-by-line_2022}; the full table is available in machine-readable form at CDS. The \Bl\ measurements of Table~\ref{tab:bl} were derived from the SPIRou/CFHT Stokes~$V$ LSD profiles, and the elemental abundances of Table~\ref{tab:abundances} were retrieved from the merged \tyc\ spectrum with the framework of \citet{jahandar_comprehensive_2024, jahandar_chemical_2025}.

\begin{table}[!t]
  \caption{Longitudinal magnetic field \Bl\ measurements of \tyc\ from SPIRou/CFHT
           Stokes~$V$ LSD profiles.}
  \label{tab:bl}
  \centering
  \begin{tabular}{lrr}
    \toprule
    \toprule
    $t_{\rm BJD} - 2\,400\,000$ & \Bl & $\sigma_{\Bl}$ \\
    (d) & \multicolumn{2}{c}{(G)} \\
    \midrule
    58745.7756 &   7.05 &  4.67 \\
    58746.9528 &  $-$0.77 & 10.91 \\
    58750.8127 &   5.68 &  4.75 \\
    58751.9052 &  11.20 &  4.54 \\
    \multicolumn{3}{c}{$\cdots$} \\
    \bottomrule
  \end{tabular}
  \tablefoot{$t_{\rm BJD}$ is the barycentric Julian date. The complete table (\nStokesVkept\ rows) is available at the CDS.}
\end{table}

\begin{table*}[!ht]
\centering
\caption{Elemental abundances of \tyc.}
\label{tab:abundances}
\setlength{\tabcolsep}{5pt}
\begin{tabular}{l cc cc cc c}
\toprule
\toprule
 & \multicolumn{2}{c}{$T_{\rm eff}=3520\,{\rm K}$}
 & \multicolumn{2}{c}{$T_{\rm eff}=3460\,{\rm K}$}
 & \multicolumn{2}{c}{$T_{\rm eff}=3580\,{\rm K}$}
 & \\
\cmidrule(lr){2-3}\cmidrule(lr){4-5}\cmidrule(lr){6-7}
$[\mathrm{X/H}]$
  & Abundance & $N_{\rm lines}$
  & Abundance & $N_{\rm lines}$
  & Abundance & $N_{\rm lines}$
  & Result\tablefootmark{a} \\
\midrule
Fe\,\textsc{i} & $+0.24\pm0.05$ & 29 & $+0.29\pm0.06$ & 28 & $+0.18\pm0.06$ & 27 & $+0.24\pm0.08$ \\
Mg\,\textsc{i} & $+0.18\pm0.06$ & 6  & $+0.34\pm0.06$ &  6 & $+0.06\pm0.05$ & 6  & $+0.18\pm0.15$  \\
Si\,\textsc{i} & $-0.47\pm0.07$ & 3  & $+0.08\pm0.24$ &  3 & $-0.72\pm0.24$ & 4  & $-0.47\pm0.41$ \\
Ca\,\textsc{i} & $+0.25\pm0.11$ & 7  & $+0.28\pm0.13$ &  7 & $+0.20\pm0.11$ & 7  & $+0.25\pm0.12$ \\
Ti\,\textsc{i} & $+0.07\pm0.07$ & 23 & $+0.16\pm0.08$ &  21 & $+0.04\pm0.06$ & 24 & $+0.07\pm0.09$ \\
Al\,\textsc{i} & $+0.48\pm0.24$ & 4  & $+0.58\pm0.24$ &  4 & $+0.40\pm0.22$ & 4  & $+0.48\pm0.26$ \\
Na\,\textsc{i} & $+0.58\pm0.11$ & 4  & $+0.65\pm0.12$ &  4 & $+0.57\pm0.12$ &  4 & $+0.58\pm0.12$ \\
C\,\textsc{i}  & $+0.28\pm0.28$ & 3  & $+0.36\pm0.26$ &  4 & $+0.21\pm0.25$ &  3 & $+0.28\pm0.29$ \\
K\,\textsc{i}  & $+0.46\pm0.16$ & 4  & $+0.45\pm0.18$ &  4 & $+0.47\pm0.13$ &  4 & $+0.46\pm0.16$ \\
OH             & $-0.41\pm0.03$ & 31 & $-0.39\pm0.04$ & 29 & $-0.45\pm0.04$ & 29 & $-0.41\pm0.04$ \\
\midrule
$[\mathrm{M/H}]$ & \multicolumn{6}{c}{} & \AaMHspirou \\
\bottomrule
\end{tabular}
\tablefoot{All abundance values are molar and relative to solar.
\tablefoottext{a}{Results combining all three $T_{\rm eff}$ grid points, so that the temperature uncertainty enters the error budget. The grid points follow the value derived from the SPIRou template spectrum, $T_\mathrm{eff}=\AaTeffSpec$.}}
\end{table*}

\begin{table}[!t]
    \caption{Radial velocities and temperature differential indicator for
                     \tyc\ from SPIRou/CFHT.}
    \label{tab:rv_dtemp}
    \centering
    \begin{tabular}{lrrrr}
        \toprule
        \toprule
        $t_{\rm BJD} - 2\,400\,000$ & $v_r$ & $\sigma_{v_r}$ & $\Delta T_{3500}$ & $\sigma_{\Delta T_{3500}}$ \\
        (d) & \multicolumn{2}{c}{(\si{\metre\per\second})} & \multicolumn{2}{c}{(K)} \\
         \midrule
        58744.8785 & 13004.76 & 3.00 & 0.3841 & 0.8125 \\
        58744.8795 & 13003.54 & 2.99 & $-$1.2272 & 0.8135 \\
        58744.8805 & 13009.03 & 2.97 & $-$0.4503 & 0.8107 \\
        58744.8816 & 13003.61 & 3.04 & $-$0.4001 & 0.8238 \\
        \multicolumn{5}{c}{$\cdots$} \\
        \bottomrule
    \end{tabular}
    \tablefoot{$t_{\rm BJD}$ is the barycentric Julian date. The complete table (\nRVframes\ rows) is available at the CDS.}
\end{table}

\section{Broadband IR colours of the unresolved system}\label{sec:ir_excess_appendix}

The NIR excess discussed in Sect.~\ref{sec:excess} is based on GNIRS photometry that resolves \innerwd\ from \tyc. As an independent check, we compared the unresolved (Aa+Ab blended) spectral energy distribution (SED) of the system, as it appears in wide-field surveys, to the field-star population of \citet{kirkpatrick_initial_2024}, the same volume-complete 20-pc census used elsewhere in this paper (Sect.~\ref{sec:excess}). Based on the resolved \AlopekeH/GNIRS contrasts (Sects.~\ref{sec:alopeke} and \ref{sec:gnirs}), \innerwd\ is $\sim$6.7--7.0\,mag fainter than \tyc\ in $J$, $H$, and $K_s$, i.e., it contributes at the sub-percent level to the system's flux in these bands, well below the photometric uncertainties used here; the unresolved photometry compared below is therefore, for practical purposes, that of \tyc\ alone.

We restricted the comparison sample to individual objects (including components of multiple systems) with a valid trigonometric parallax and a spectral type of M3.0--M4.9, bracketing \tyc's own M3.9V classification \citep{fouque_spirou_2018}; this yielded \IRexcessNmatched\ comparison stars. Because flux falls steeply with spectral type across this range, we worked throughout in colour space: for each band we took the pool's median offset from \textit{Gaia} $G$ (3$\sigma$-clipped), and combined it with \tyc's own observed $G$-band flux to predict what a typical M3.0--M4.9 dwarf, scaled to \tyc's own optical flux, should look like at every other wavelength.

\begin{figure}[!t]
    \centering
    \includegraphics[width=\linewidth]{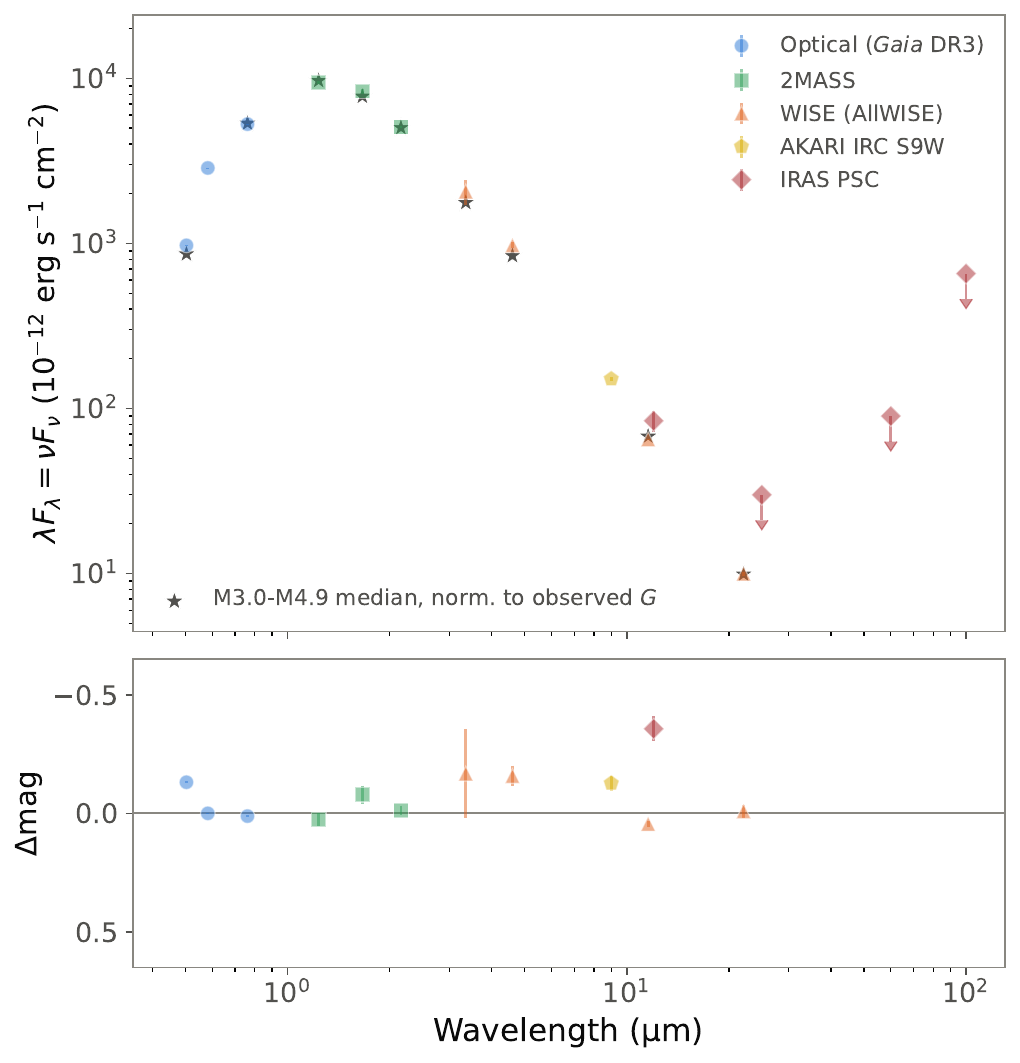}
    \caption{SED of the unresolved \welo\ Aa+Ab system (coloured points, by survey) compared to the \citet{kirkpatrick_initial_2024} 20-pc census, restricted to objects with a spectral type of M3.0--M4.9, including components of multiple systems (\IRexcessNmatched\ objects; grey stars), normalised to the system's own observed \textit{Gaia} $G$ magnitude. Top: $\lambda F_\lambda = \nu F_\nu$ versus wavelength; IRAS upper limits (25--100\,$\mu$m) are shown with downward arrows. Bottom: $\Delta$mag (observed minus pool-predicted) at each wavelength, on an inverted axis, so that a point brighter than predicted lies upwards, with AKARI and the IRAS 12\,$\mu$m detection referenced to a log($\lambda$)--log($F$) interpolation between the pool-predicted $W1$--$W4$ fluxes.\label{fig:sed_appendix}}
\end{figure}

Figure~\ref{fig:sed_appendix} shows the resulting SED (top) and the deviation from the pool prediction, $\Delta$mag, at each wavelength (bottom). The observed SED is closely consistent with the M3.0--M4.9 pool from the optical through the MIR: all bands other than WISE $W1$/$W2$ and the IRAS \SI{12}{\micro\metre} point discussed below land within $\sim$\IRexcessTypicalOffset\,mag of the pool-predicted value, including 2MASS $J$, $H$, $K_s$ and WISE $W3$/$W4$ (\IRexcessWthreeOffset\,mag). \tyc\ runs $\sim$\IRexcessWoneOffset\,mag brighter than the pool median at $W1$ and $\sim$\IRexcessWtwoOffset\,mag at $W2$, a modest, two-band-consistent effect that is small relative to the pool's own band-to-band scatter (\IRexcessPoolScatter\,mag) and far below what a genuine warm-dust or companion-driven MIR excess would produce; it echoes, in size and wavelength range, the tentative NIR excess reported towards \innerwd\ in Sect.~\ref{sec:excess}. The AKARI/IRC 9\,$\mu$m detection, referenced to the same pool-predicted interpolation, is consistent with this same modest excess ($\sim$\IRexcessAKARIoffset\,mag).

The IRAS Point Source Catalog 12\,$\mu$m point is the only discrepant point: it sits $\sim$\IRexcessIRASoffset\,mag ($\sim$\IRexcessIRASratio\ in flux) brighter than the pool-predicted flux at that wavelength. This is surprising because AllWISE $W3$, an almost identical bandpass (pivot wavelength 11.56\,$\mu$m against IRAS's 12\,$\mu$m), is itself unremarkable ($\sim$\IRexcessWthreeOffset\,mag from the pool prediction), unlike the marginal excess noted above at $W1$/$W2$; since $W3$ shows no such effect while the archival IRAS point stands out sharply, we consider it far more likely that the IRAS Point Source Catalog measurement itself is compromised, whether by the lower angular resolution and calibration of the survey or by genuine MIR variability of the system between the IRAS and WISE epochs, than that it reflects real circumstellar emission; we cannot distinguish between these possibilities with the data in hand and leave the question to future monitoring. The 25--100\,$\mu$m IRAS points remain upper limits well above any plausible photospheric level.

\end{appendix}

\end{document}